\documentclass[12pt]{article}
\usepackage{amssymb,amsmath}
\usepackage{latexsym}
\usepackage{graphicx}
\usepackage{color}
\usepackage{datetime}
\usepackage{cite}
\usepackage[
      colorlinks=false,
      linkcolor=darkblue,  
      urlcolor=blue,    
      filecolor=blue,     
      citecolor=red,
linktocpage=true,
      pdfstartview=FitV,
      bookmarksopen=true    
      ]{hyperref}

\usepackage{psfrag}

\definecolor{cardinal}{rgb}{0.6,0,0}
\definecolor{darkgreen}{rgb}{0,0.5,0}
\definecolor{golden}{rgb}{0.92, 0.7, 0}
\definecolor{midnight}{rgb}{0, 0, 0.5}
\definecolor{darkblue}{rgb}{0.2, 0, 0.8}

\def\coeff#1#2{\relax{\textstyle {#1 \over #2}}\displaystyle}

\def\IR{\mathbb{R}}

\def\cH{{\cal H}}

\def\cN{{\cal N}}

\def\cR{{\cal R}}

\def\cO{{\cal O}}

\def\bbR{\mathbb{R}}

\def\bbZ{\mathbb{Z}}

\def\ket#1{|{#1}\rangle}

\begin{document}

\begin{titlepage}

\begin{flushright}
IPhT-T14/072\\
YITP-14-36
\end{flushright}

\centerline{\Large \bf Black-Hole Entropy from Supergravity Superstrata States}
\date{\today}

\bigskip
\centerline{{\bf Iosif Bena$^1$, Masaki Shigemori$^{2,3}$ and Nicholas P. Warner$^{4,1}$}}
\bigskip
\centerline{$^1$ Institut de Physique Th\'eorique, }
\centerline{CEA Saclay, F-91191 Gif sur Yvette, France}
\bigskip
\centerline{$^2$ Yukawa Institute for Theoretical Physics, Kyoto University}
\centerline{Kitashirakawa Oiwakecho, Sakyo-ku, Kyoto 606-8502 Japan}
\bigskip
\centerline{$^3$ Hakubi Center, Kyoto University}
\centerline{Yoshida-Ushinomiya-cho, Sakyo-ku, Kyoto 606-8501, Japan}
\bigskip
\centerline{$^4$ Department of Physics and Astronomy,}
\centerline{University of Southern California,} \centerline{Los
Angeles, CA 90089, USA}
\bigskip

\begin{abstract}
\noindent
There are, by now, several arguments that superstrata, which represent
D1-D5-P bound states that depend upon arbitrary functions of \emph{two
variables} and that preserve four supersymmetries, exist in string
theory, and that their gravitational back-reaction results in smooth
horizonless solutions.
In this paper we examine the shape and density modes of the superstratum
and give strong evidence that the back-reacted supergravity solution
allows for fluctuation modes whose quantization reproduces the
entropy growth of black holes as a function of the charges.
In particular, we argue that the shape modes of the superstratum that
lie purely within the non-compact space-time directions account for at
least $1/\sqrt{6}$ of the entropy of the D1-D5-P black hole and propose
a way in which the rest of the entropy could be captured by superstratum
fluctuations.
We complete the picture by conjecturing a relationship between bound
states of multiple superstrata and momentum excitations of different
twisted sectors of the dual CFT\@.

\end{abstract}

\end{titlepage}



\section{Introduction}

\subsection{Black-hole microstate structure}

The prototypical example of a string theory black hole whose entropy can
be accounted for microscopically is the D1-D5-P black hole. If one
considers the various ways in which a combination of $N_1$ D1 and $N_5$
D5 branes can carry $N_P$ units of momentum (in the regime of parameters
where the back-reaction of these branes is not important and the
physical picture of the momentum-carrying excitations is clear), one
finds that the corresponding entropy is given by $2 \pi \sqrt{N_1 N_5
N_P}$, which exactly matches the Bekenstein-Hawking entropy of the black
hole that these branes form in the regime of parameters where their
back-reaction is important. Since the original work of \cite{Sen:1995in,
Strominger:1996sh}, such entropy-matching calculations have been
extended to many other families of supersymmetric, or merely extremal black
holes,  and even near-extremal black holes.  The matching of the entropies has proven remarkably
successful.

In 1996, the perturbative counting of black-hole microstates at
vanishing string coupling in \cite{Strominger:1996sh} represented the
first real progress on the microstate problem in many years. However,
this work opened up a whole new set of questions. In particular, it
remained to understand how \emph{one} particular black-hole microstate
manifests itself in the finite-coupling regime in which the classical
black-hole solution exists and has a large horizon area.  For a long
time it had been thought that all the microstates at weak coupling
develop a horizon and are indistinguishable from the classical
black-hole solution (except perhaps in a Planck-size region around the
singularity) \cite{Horowitz:1996nw, Damour:1999aw}. This intuition was
challenged by the construction of several families of fully back-reacted
solutions that have the same charges and mass as the black hole, but
differ from the classical black-hole solution at the scale of the
horizon and, in particular, are smooth and horizonless
\cite{Bena:2005va, Berglund:2005vb}. Such solutions are called
``microstate geometries,'' because, via the AdS/CFT correspondence, one
can map them onto states of the dual CFT\@.
However, despite having many properties indicating that they belong to
the typical sector of the black-hole microstates, these solutions have
an entropy that is parametrically lower than the black-hole entropy \cite{deBoer:2009un},
which is presumably related to the fact that these solutions have a lot
of symmetry.

If one is to try to reproduce the black hole entropy from supergravity
one should therefore find solutions with less symmetry, and the first
step in this direction was the construction of three-charge solutions
that contain a wiggly supertube \cite{Mateos:2001qs}.  These solutions
are parametrized by an arbitrary continuous function and hence can have
an infinite number of continuous parameters \cite{Bena:2010gg}. The
entropy of these solutions grows with the charges as $N^{5/4}$
\cite{Bena:2010gg}, which is more than all other known supergravity
solutions, but is still less than the black hole entropy growth,
$N^{3/2}$. In \cite{Bena:2011uw} we have furthermore argued that if one
relaxes one more symmetry one can construct smooth horizonless
\emph{superstratum} solutions that depend on arbitrary continuous
functions of {\em two} variables, and it is the purpose of this paper to
argue that the perturbative semi-classical quantization of superstrata
yields a black-hole-like entropy growth, and that in the fully
back-reacted regime all the three-charge black-hole entropy might be
reproduced by space-time fluctuation modes of the superstrata.

In parallel with our efforts, there have also been several
relatively-recent developments that support this general approach. First
amongst these is Mathur's tightening \cite{Mathur:2009hf, Mathur:2012np,
Mathur:2012dxa} of Hawking's result to show that information can only be
recovered if there are  $\cO(1)$ corrections to the semi-classical
physics outside black holes. That is, in order to solve the information
problem, we need to make some $\cO(1)$ changes at the horizon scale.
This discussion can be taken to a new level by asking whether these
changes result in a firewall for an incoming observer, as argued by
\cite{Braunstein:2009my, Almheiri:2012rt, Mathur:2012jk,
Susskind:2012rm, Bena:2012zi, Susskind:2012uw, Avery:2012tf,
Avery:2013exa, Almheiri:2013hfa, Verlinde:2013uja, Verlinde:2013uja,
Mathur:2013gua} or rather whether the quantum superposition of these
states can result in a smooth infall experience for macroscopic
infalling observers \cite{Mathur:2010kx, Mathur:2011wg, Mathur:2012jk,
Mathur:2012zp}. However, finding a mechanism that can support such
$\cO(1)$ changes in the structure at the horizon scale is notoriously
difficult -- essentially because the horizon is null, any massive object
must fall in, while any massless wave packet will dilute to nothing
after several horizon-crossing times. The {\em only} time-independent
way to support such a structure within supergravity is to place magnetic
fluxes on topologically non-trivial cycles \cite{Gibbons:2013tqa,
Haas:2014spa}, and this is precisely the mechanism that underpins all
the known BPS \cite{Bena:2005va, Berglund:2005vb, Bena:2007kg} and
near-extremal \cite{Bena:2011fc, Bena:2012zi} microstate
geometries. Furthermore, as we have argued in \cite{Bena:2013dka}, this
mechanism extrapolates well beyond the regime of validity of
supergravity, and can manifest itself either via brane polarization
\cite{Myers:1999ps} or via non-Abelian effects.

As explained in \cite{Bena:2013dka}, there are  two separate
issues that one must address in order to understand the microstate
structure of black holes and the effect that this structure has at the horizon
scale. The first is how one can make changes at the horizon scale and 
we now know \cite{Gibbons:2013tqa} that the geometric
transition discovered in five dimensions \cite{Bena:2005va,
Berglund:2005vb} provides the \emph{only} way to replace the horizon with
horizonless time-independent structure thereby making the $\cO(1)$ corrections.   
Such  geometric transitions will therefore  be an essential part of any
string-based resolution of black-holes. The microstate structure itself,
whatever its ultimate form, can then be supported by the ``canvas''
provided the geometric transition to large microstate geometries. 

The second issue is to determine the extent to which this microstate structure can be
captured by semi-classical geometries.  This paper will advance the
latter goal by arguing that there is indeed a class of microstate
geometries, called superstrata, that can achieve the second goal at
least with sufficient fidelity to obtain the correct charge-dependence
of the BPS black-hole entropy.

\subsection{Superstrata}
\label{ss:superstrata}

The \emph{superstratum} is a smooth, horizonless soliton (a microstate
geometry) that is $\frac{1}{8}$-BPS (preserving 4 supersymmetries),
depends on several arbitrary functions of two variables and has the same
charges as the D1-D5-P black hole. The existence of this object was
conjectured in \cite{Bena:2011uw} (building on earlier work in
\cite{deBoer:2010ud}) by arguing that a certain combination of branes,
Kaluza-Klein monopoles (KKM's) and momentum preserves the same
supersymmetries as the D1-D5-P black hole irrespective of its
orientation, and hence one can glue these branes into a supersymmetric
configuration that depends on functions of two variables. Furthermore,
since the superstratum locally resembles a D1-D5 supertube with a KKM
dipole charge, the fully back-reacted superstratum solution should be
smooth and hence be a microstate geometry.  Even though there is not yet
an explicit construction of a generic fully back-reacted superstratum,
one can find further evidence for their existence by analyzing string
emission from the D1-D5-P system \cite{Giusto:2012jx, Giusto:2013bda,
Shigemori:2013lta}, or by constructing supergravity solutions that
depend of two different functions of two different variables
\cite{Niehoff:2013kia}, which could be thought of as limits of the more
general superstratum solution.

There are several ways by which one might realize the construction of a
superstratum.  The first way is via a double supertube transition
\cite{deBoer:2010ud, Bena:2011uw, deBoer:2012ma}: one combines the D1
branes with some momentum to give a D1-P supertube (D1's with traveling
waves on them) and, at the same time, one combines some D5 branes with
some momentum to obtain a D5-P supertube (D5's with traveling waves on
them).  One must do this in such a manner that the D1-profile lies
entirely within the D5-profile.  Next one ``executes'' a second
supertube transition by locally puffing out the D1-D5 system using a
Kaluza-Klein monopole and the result is a D1-D5-P bound state.  Since
supertube transitions  give the configuration an arbitrary
profile and the second transition can, in principle, be done independently and locally
on each D1-D5 segment, it seems plausible
\cite{Bena:2011uw} that two supertube transitions could give rise to a
smooth superstratum solution that can be parametrized by functions of
two variables.

The second way to think of a superstratum is to begin with a D1-D5
supertube with KKM dipole charge (parametrized by several arbitrary
functions of one variable) and start adding momentum to it. Again, for
each original configuration, given by the Lunin-Mathur geometry
\cite{Lunin:2001jy, Lunin:2002iz, Emparan:2001ux} one expects to be able
to add a general wave profile along the common D1-D5 direction, and
hence to obtain a configuration that depends on functions of two
variables. Thus, every mode of the original D1-D5 supertube will act as
a momentum carrier, and therefore the number of carriers over which one
can distribute a given momentum is the number of modes of the D1-D5
supertube.  This suggests that such excitations should describe a moduli
space of D1-D5 supertubes, and each such modulus should be able to
carry momentum.

A third perspective on superstrata comes from the fact that they
describe bubbled microstate geometries.  Indeed, the single, circular,
unexcited superstratum is identical to a D1-D5 supertube geometry and
this geometry, in the near-tube limit, is, up to orbifolding, the
maximally-symmetric geometry \emph{global} $AdS_3 \times S^3$
\cite{Lunin:2001jy}.  More generally, multiple superstrata are expected
to describe geometries with topological $3$-cycles held up by
cohomological fluxes.  Changing the shapes of the superstrata
corresponds to changing the shapes of these cycles and letting these
shape changes depend upon the compact circle in $AdS_3$.  On a single
superstratum, the modes transform under the isometries $SL(2,\IR)_L
\times SL(2,\IR)_R$ $\times SU(2)_L \times SU(2)_R$.  If the structure
is to carry momentum then supersymmetry requires that this momentum be
either purely left-moving or purely right-moving and so BPS fluctuations
can only excite half the modes.  As we will discuss in Section
\ref{MomStates}, within the D1-D5 CFT, the left-moving excitations in the space-time
directions are correlated with fermionic excitations that only carry $SU(2)_L$ 
quantum numbers\footnote{This observation also
has interesting implications for future work: near-BPS and non-BPS
solutions have long been obtained by exciting both left-moving and
right-moving momentum \cite{Callan:1996dv, Horowitz:1996fn,
Breckenridge:1996sn, Horowitz:1996ay} and so we expect generic shape
fluctuations to be a natural way to access such non-BPS
solutions.}.  It is this that places restrictions on the BPS modes and thus
upon the perturbative shape fluctuations. This perturbative approach to superstrata has been
developed in \cite{Giusto:2012jx, Giusto:2013bda} and very simple, restricted 
classes of fully back-reacted solutions were described in \cite{Niehoff:2013kia}.

\subsection{Representing black hole microstates with superstrata}

The problem with the quantization of the superstratum is that we do not know its action and so we cannot start from first principles and quantize.  On the other hand we do know the perturbative description of the D1-D5-P microstates that give the black-hole entropy and we know the field theory dual of the $AdS_3 \times S^3$ solution corresponding to the unexcited superstratum.  From these observations we can ``reverse engineer'' precisely which states of the superstratum will be visible within  supergravity.   Our ultimate goal  is to argue that the modes of the D1-D5-P system will, in supergravity, give rise to geometric modes whose semi-classical quantization will reproduce the exact black-hole entropy:
\begin{equation}
S ~=~ 2 \pi \sqrt{ N_1 N_5 N_P} \label{count1} \,. 
\end{equation}

We will, however, start far more conservatively with what we believe can
be substantiated with a high level of confidence, namely, that the
semi-classical quantization of the space-time shape modes of a
\emph{single} superstratum can lead to an entropy count of, at least,
\begin{equation}
S ~=~ 2 \pi \sqrt{\frac{1}{6} N_1 N_5 N_P} \,. \label{count2} 
\end{equation}
This differs from \eqref{count1} by a factor of $ \frac{1}{\sqrt{6}}$
because, as we will discuss, the perturbative space-time shape modes of
a single superstratum must involve only one sixth of the complete set of
perturbative BPS modes.  More precisely, these BPS space-time shape
modes describe a sector of the CFT with central charge $c= N_1 N_5$
corresponding to half of the bosonized fermions in the D1-D5 CFT\@.  The
remaining part of the CFT, with central charge $c=5 N_1 N_5$, arises
from the other half of the bosonized fermions and the original bosonic
excitations of the D1-D5 CFT\@.  These states correspond to corrections
to the internal metric and fields on the $T^4$ upon which the D5 branes
are compactified.  We will examine the extent to which this ``other
five-sixths'' of the BPS states will be visible within supergravity and
 argue that in the fully back-reacted regime the modes that contain internal torus fluctuations will have an energy gap that is parametrically larger than that of the typical black hole microstates. We  suggest that these internal torus modes will be ``pushed on the Coulomb branch'' and will become visible as transverse supergravity modes of the superstratum solution. 
 
 The important point here is that, whatever the ultimate status of the internal $T^4$ excitations,  the  arguments based upon group theory and perturbation theory allow us to assert with considerable confidence that the shape modes of a single  superstratum can, at least, recover the correct entropy growth $S\sim \sqrt{N_1 N_5 N_P}$ as a function of $N_1 N_5 N_P$. 

It is also possible to estimate the entropy of superstrata by starting
from the original argument \cite{Bena:2011uw} that they can be
constructed as momentum-carrying fluctuations of the D1-D5
supertube. This construction appears to allow {\em all} the shape modes
of the supertube to be promoted to momentum carriers\footnote{This also
agrees with the physics of certain explicit solutions that can be
thought of as singular limits of the superstratum solution
\cite{Bena:2011dd, Niehoff:2012wu}.}. We will argue in Section \ref{5/6} that the
dimension of the moduli space of these shape modes is $4N_1N_5$, which
would imply that the entropy of a superstratum will come from
distributing $N_P$ units of momentum over $4 N_1 N_5$ bosonic carriers
and their fermionic superpartners, and this would reproduce exactly the
black-hole entropy (\ref{count1}). This construction appears to be at
odds with the perturbative analysis that gives the entropy
(\ref{count2}).  It is possible that the $4 N_1 N_5$ shape modes are not
independent and unobstructed moduli.  It is also quite possible, as we
will also discuss in Section \ref{5/6}, that the extra shape modes that
go beyond the perturbative analysis of Section \ref{MomStates} will only
emerge in the fully back-reacted superstratum solution.  We therefore
hope that an complete and explicit superstratum solution will clarify
whether the space-time modes of the superstrata will reproduce all the
black-hole entropy or only $1\over \sqrt{6}$ of it.

In formulating the entropy-counting arguments above we have taken it as
given that adding momentum charge to a BPS system of branes will always
lead to transverse shape modes once the supergravity back-reaction is
included.  We will also assume the converse: semi-classical quantization
of such supergravity shape modes will recover a full description of the
Hilbert space of the original perturbative momentum modes.  This is
certainly true of the F1-P system, since this is simply the quantization
of the fundamental string \cite{Dabholkar:1995nc} and it is also true of
momentum modes on many systems of branes.  We do not believe that there
is much danger in assuming that this is a universal
result\footnote{Strictly speaking, this must hold for the momentum added
to the unique ground state of the system and does not apply to the
momentum carried by the ground state itself.  We are always concerned
with the former.  For example, a straight supertube \cite{Mateos:2001qs}
carries a fixed amount of angular (longitudinal) momentum coming from
the crossing of electric and magnetic worldvolume fluxes.  However, any
change in the momentum on top of that leads to transverse fluctuation of
the supertube shape and of the back-reacted supergravity solution
\cite{Palmer:2004gu, Bak:2004rj, Bak:2004kz}.\label{ftnt:transv_fluct}}.

There are two frequently-expressed concerns about any program, as the
one advanced here, that involves obtaining the black-hole entropy by
counting supergravity solutions.  The first is that classical
supergravity modes only correspond to coherent quantum states and that
the states that contribute to the entropy cannot be geometric.  The
second is that it is possible that the
fluctuations that contribute primarily to the entropy may have very
small scales, and hence the corresponding solutions will have structure
below the Planck scale and will not be therefore correctly described by
supergravity.

The first concern might equally be raised as an objection to considering the vibrational motion of a diatomic molecule to be that of a spring.  Obviously this is a dramatic classical simplification of a complex quantum system and the real motions of a diatomic molecule are intrinsically quantum phenomena.  However, approximating the chemical bond by a classical harmonic oscillator and semi-classically quantizing this oscillator gives an excellent description of the quantum states and the vibrational spectrum because the ``spring'' isolates the essential physical degrees of freedom that govern the system.  It is in this spirit that we believe that microstate geometries and their semi-classical quantization will describe sufficiently many microstates of black holes and give a valuable description of their thermodynamics:  While the quantum mechanical states of a black hole are manifestly not geometric, and only very few of them have classical descriptions, the important insight coming from microstate geometries is that this allows us to identify the \emph{degrees of freedom at strong coupling} that need to be quantized in order to capture the essential underlying physics of the black-hole microstates.

The second concern is more serious in that the entropy might be coming
primarily from a sector in which the supergravity approximation is
failing.  There are two reasonable ways around this issue.  First, we
know that exactly the same issue arises in other instances of adding
momentum modes to branes, as with the fundamental string, and yet
there is no problem with the semi-classical quantization of states.  The
reason why there is no difficulty is precisely because such states are
based upon well-understood systems of objects that make sense in string
theory.  Thus the easiest answer to the second concern is that we may
ultimately have to broaden the scope of the semi-classical quantization
and go beyond smooth microstate geometries, whose scales, by definition,
lie comfortably above the Planck length, and include \emph{microstate
solutions}. The latter are defined \cite{Bena:2013dka} to be
horizonless, physical limits of smooth geometries that have the same
mass, charge and angular momentum as a given black hole, but can have
singularities that either correspond to fundamental (${1\over 2}$-BPS)
D-brane sources or can be patch-wise dualized into a smooth solution.

It is also possible that smooth microstate geometries will resolve these issues without needing to introduce stringy singularities. Indeed, one important realization in the study of microstate geometries was that if one wants to construct a solution that has the same charges as a five-dimensional three-charge black hole with a macroscopically-large horizon area, one must use \emph{scaling solutions} \cite{Bena:2006kb, Denef:2007vg, Bena:2007qc}. In these solutions the size of the bubbles appears to shrink to zero size from the perspective of the metric of the auxiliary four-dimensional base-space that is used to construct the solutions, but, in fact, the bubbles remain finite once the supergravity back-reaction is taken into account. In the scaling limit, these bubbles descend down a very long $AdS$ throat that resembles, more and more, that of the corresponding black hole. Hence, it is possible that adding a third charge to what appear to be very stringy two-charge microstates will expand the physical length scales and result in smooth fluctuating solutions at the bottom of a very long throat.

\subsection{The present approach}

Returning to our main goal, we wish to describe the detailed structure
of the semi-classical superstratum in terms of the D1-D5 CFT\@.  We
therefore begin in Section \ref{Sect:D1D5CFT} by reviewing the D1-D5 CFT
and in Section \ref{Sect:2Chg} we describe the two-charge ($\frac{1}{4}$-BPS)
states of the D1-D5 system and how they correspond to
supertube profiles.  In Section \ref{MomStates} we add momentum to the
system and relate the three-charge ($\frac{1}{8}$-BPS) states to
profiles of the superstratum. We initially adopt a rather
conservative approach by focussing on the details of the microstate
structure that we are confident can be reproduced by quantizing the
supergravity modes. In particular, we focus on the space-time shape
modes of the superstratum and how they can be matched to perturbative
modes of a particular sector of the D1-D5 CFT\@. This allows us to reproduce the correct charge growth of the black-hole entropy, albeit with a smaller overall coefficient.
In Section \ref{5/6} we adopt a less conservative view of the possible
modes that a superstratum can have, which is closer to the original
arguments for the existence of superstrata \cite{Bena:2011uw} and to the
physics of certain singular limits of superstratum solutions
\cite{Bena:2011dd, Niehoff:2012wu}. This allows us to use a counting argument similar to that of Maldacena, Strominger and Witten
\cite{Maldacena:1997de} to reproduce exactly the entropy of the three-charge black hole, and to
obtain the correct overall coefficient as well. We then discuss several
ways in which the liberal and conservative approaches to superstrata can
be reconciled, and in particular we suggest in Section \ref{Sect:Multi}
that bound states of multiple superstrata may be a key ingredient in
relating all the states of the CFT to bulk supergravity
solutions.  Section \ref{Sect:Concl} contains our concluding
remarks.

\section{The D1-D5 CFT and the ``visible'' sector}
\label{Sect:D1D5CFT}

The easiest way to quantize the two-charge system is in the F1-P frame
where the states are simply those of the perturbative string.  However,
for the superstratum, we are going to need the detailed description in
the D1-D5 duality frame where there are $N_5$ D5 branes wrapped on $T^4
\times S^1$ and $N_1$ D1 branes wrapped on the common $S^1$. Let $R$ be
the radius of the $S^1$ and $v$ the corresponding coordinate.
For fixed $v$, the moduli space of the configurations is the same as
that of $N_1$ D0 branes inside $N_5$ D4 branes and so it may be
identified with the moduli space of $N_1$ instanton sector of $SU(N_5)$
Yang-Mills.  The dimension of this moduli space is $4 N_1 N_5$.
These moduli can be made into functions of $v$ and thus, in the
perturbative regime, one has a CFT with $4 N_1 N_5$ bosons on this
$S^1$.  However, the D1-D5 system has 8 supersymmetries, which extend
the CFT to an $\cN=(4,4)$ SCFT\@. There are thus $8 N_1 N_5$ free
fermions that split into $4 N_1 N_5$ left-movers and $4 N_1 N_5$
right-movers.\footnote{For more details on the D1-D5 CFT, see, for example, 
\cite{Aharony:1999ti, David:2002wn, Avery:2010qw, Giusto:2013bda}.}
 
To be more precise, the underlying field theory is the $\cN=(4,4)$
superconformal sigma model whose target space is the orbifold,
$(T^4)^N/S_N$, where $N\equiv N_1 N_5$ and $S_N$ is the permutation
group on $N$ elements\footnote{This is the description of the CFT at the
free orbifold point.}. There are thus $4N$ free bosons and $4N$ free
fermions.  Following \cite{Avery:2010qw, Giusto:2013bda} the bosons will
be labeled, $X^{\dot{A} A}_{(r)} (z,\bar{z})$, where $r =1, \dots, N$,
is the copy index of the $T^4$ and $A,\dot{A}=1,2$ are spinorial
indices for the $SO(4)_I=SU(2)_1\times SU(2)_2$ of the tangent space of
$T^4$.  The left-moving and right-moving fermions, $\psi_{(r)}^{\alpha
\dot{A}}(z)$ and $\tilde\psi_{(r)}^{\dot{\alpha} \dot{A}}(\bar{z})$ with
$\alpha,\dot{\alpha}=\pm$, transform as doublets of
fixed helicity on the $T^4$ and as doublets of different helicities
under the $\cR$-symmetry, $SO(4)_\cR = SU(2)_L \times SU(2)_R$. Note
that the fermions transforming in the $({\bf 2},{\bf 1})$ and $({\bf
1},{\bf 2})$ of the $\cR$-symmetry are left-moving and right-moving,
respectively.  The $T^4$ is, of course, the compactification manifold of
the D5's and, as usual in theories on D-branes, the $\cR$-symmetry is
generated by rotations in the (non-compact) spatial directions
transverse to all the branes, that is, in the space-time directions.

In the fully back-reacted D1-D5 geometry, the near-brane limit is \emph{global} $AdS_3
\times S^3\times T^4$ and the symmetry outside the $T^4$ is $ SL(2,\IR)_L
\times SU(2)_L  \times  SL(2,\IR)_R \times SU(2)_R$.  These symmetries
correspond to the left-moving and right-moving (finite) conformal invariance
and $\cR$-symmetry via the holographic duality.

By construction, the excitations of the bosons, $X^{\dot{A} A}_{(r)}$,
only involve motions in the compactified ($T^4$) directions, whereas the
fermionic excitations carry polarizations ($\cR$-charge) that \emph{are}
visible within the six-dimensional space-time.  To understand what portion of the
fermion Hilbert space is visible from the space-time, it is convenient
to bosonize the fermions by defining the currents
\begin{align}
J_{(r)}^{\alpha \beta}(z)  & ~\equiv~  \frac{1}{2} \psi_{(r)}^{\alpha \dot{A}}(z)  \, \epsilon_{\dot{A}\dot{B}} \,  \psi_{(r)}^{\beta \dot{B}}(z)  \,,  \qquad \tilde J_{(r)}^{\dot{\alpha} \dot{\beta}}(\bar{z})   ~\equiv~  \frac{1}{2} \tilde\psi_{(r)}^{\dot{\alpha} \dot{A}}(\bar{z})  \, \epsilon_{\dot{A}\dot{B}} \,  \tilde\psi_{(r)}^{\dot{\beta} \dot{B}} (\bar{z}) \,, \label{Jdefn}\\
K_{(r)}^{\dot{A} \dot{B}} (z) & ~\equiv~  \frac{1}{2} \psi_{(r)}^{\alpha \dot{A}} (z) \, \epsilon_{\alpha \beta} \,  \psi_{(r)}^{\beta \dot{B}}(z)  \,,  \qquad \tilde K_{(r)}^{\dot{A} \dot{B}} (\bar{z})  ~\equiv~  \frac{1}{2} \tilde\psi_{(r)}^{\dot{\alpha} \dot{A}} (\bar{z}) \, \epsilon_{\dot{\alpha} \dot{\beta}} \,  \tilde\psi_{(r)}^{\dot{\alpha} \dot{A}} (\bar{z})  \,.
\end{align}
For each value of $r$, the currents $J_{(r)}^{\alpha \beta}$ and $\tilde J_{(r)}^{\dot{\alpha} \dot{\beta}}$ generate a level $1$, $SU(2) \times SU(2)$ current algebra.  Each such algebra may be viewed as being generated by a single boson. 

If one sums over $r$, the currents
\begin{equation}
J^{\alpha \beta}(z)  ~\equiv~  \sum_{r=1}^N \, J_{(r)}^{\alpha \beta}(z)   \,,  \qquad \tilde J^{\dot{\alpha} \dot{\beta}}(\bar{z})  ~\equiv~  \sum_{r=1}^N \, \tilde J_{(r)}^{\dot{\alpha} \dot{\beta}}(\bar{z})   \label{CFTUones}   \,,  
\end{equation}
generate the level $N$, $SU(2)_R \times SU(2)_L$ current algebra of the
$\cR$-symmetry.  Because of the pseudo-reality of the fermions \cite{Avery:2010qw, Giusto:2013bda}, the
standard angular momentum operators, $J^\pm$ and $J^3$, are given in
terms of the $J^{\alpha \beta}$ by:
\begin{equation}
\begin{split}
 J_L^3 &~=~ J^{1 2} ~=~ J^{2 1}    \,,  \qquad 
 J_L^+  ~=~ J^{1 1}\,,  \qquad 
 J_L^-  ~=~ J^{2 2}   \,;\\
 J_R^3 &~=~ \tilde J^{1 2} ~=~ \tilde J^{2 1}    \,,  \qquad 
 J_R^+  ~=~ \tilde J^{1 1}\,,   \qquad
 J_R^-  ~=~ \tilde J^{2 2}     \,.
\end{split}
\end{equation}

For each value of $r$, the currents $K_{(r)}^{\dot{A} \dot{B}}$ and
$\tilde K_{(r)}^{\dot{A} \dot{B}}$ also generate level $1$, $SU(2)_1$
current algebras but now purely on the $T^4$.\footnote{For the full
internal $SU(2)$ symmetry current, we must include the contribution from
the bosonic field $X^{\alpha\dot{\alpha}}$.  Note that the
internal rotational symmetry is, of course, broken by the
compactification.}  The important point is that the $J_{(r)}^{\alpha
\beta}(z)$ and $K_{(r)}^{\dot{A} \dot{B}}(z)$ are completely
``orthogonal'' sets of operators that commute with one
another\footnote{One can see this most easily by viewing the indices on
the fermions, $\psi_{(r)}^{\alpha \dot{A}}$, as transforming as a $({\bf
2},{\bf 2})$ of $SU(2)_L \times SU(2)_1$ and then the $J$'s and $K$'s
generate these two $SU(2)$'s.} and similarly for $\tilde
J_{(r)}^{\dot{\alpha} \dot{\beta}}$ and $\tilde K_{(r)}^{\dot{A}
\dot{B}}$.  Thus the $N$ $SU(2)$ current algebras generated by the
$J_{(r)}^{\alpha \beta}$ and $\tilde J_{(r)}^{\dot{\alpha} \dot{\beta}}$
involve excitations that are purely visible from the space-time with no
component of this chiral algebra creating an excitation on the torus.
Conversely, the $K_{(r)}^{\dot{A} \dot{B}}$ and $\tilde K_{(r)}^{\dot{A}
\dot{B}}$ represent the chiral algebras that are visible only from the
$T^4$ and invisible from the space-time. Thus the perturbative
excitations that are visible from the six-dimensional space-time form Hilbert spaces, $\cH_{\rm st}$, that can be
characterized by the representations of, and excitations created by, the
conformal field theory:
\begin{equation}
(SU(2)_L\times SU(2)_R)^N/S_N \label{ShapeCFT} \,,  
\end{equation}
where the $J_{(r)}$ and $\tilde J_{(r)}$ generate these level $1$
current algebras.  This theory has central charge $c = N = N_1 N_5$.  Similarly, 
the CFT that lies purely on the internal directions has $c = 5N = 5N_1
N_5$ and is generated by the bosons, $X^{A \dot{A}}$, and the currents
$K_{(r)}$ and $\tilde K_{(r)}$.  We will denote the internal Hilbert spaces  by $\cH_{\rm int}$ and think
of the states of the D1-D5 theory as being decomposed into a sums  of the products of the form
\begin{equation}
\cH~=~ \cH_{\rm st}\otimes \cH_{\rm int} \,.
\label{HilbSpaces}
\end{equation}

The back-reaction of the fermionic and bosonic modes of the D1-D5 CFT will result in shape and charge-density modes of the corresponding supergravity solution.  Conversely, we will argue, in the next section, that the semi-classical quantization of the corresponding families of BPS microstate geometries will lead to the states of the   D1-D5 CFT\@.  Indeed this is precisely what holographic field theory on $AdS_3 \times S^3$ suggests.  Moreover, because of the split into $c = N = N_1 N_5$ and $c = 5N = 5N_1 N_5$ sectors detailed above, we expect that the supergravity modes in the space-time directions alone will be enough to see a $c = N_1 N_5$ sector of the CFT while the remaining $c = 5 N_1 N_5$ sector will be visible from semi-classical quantization of internal modes of the D1-D5 system.

We now substantiate this view by revisiting the geometry and semi-classical structure of the two-charge system and argue how this will be modified via the addition of the third charge via momentum modes. 

\section{The two-charge states}
\label{Sect:2Chg}
  
The two-charge states of the D1-D5 system are the Ramond-Ramond (RR)
ground states of the CFT and preserve half the CFT supercharges, or
eight supersymmetries (note that these states are called
$\frac{1}{4}$-BPS states, relative to the 32 supercharges of type IIB
superstring before putting D-branes).  These states have angular momenta
in the range $- \frac{N}{2} \le J^3_L, J^3_R \le \frac{N}{2}$.  One can
spectrally flow these states to the NS sector to obtain chiral primary
fields and the RR ground states can viewed as being created by chiral
primaries acting on the maximally-spinning RR ground state,
$\ket{\psi_0}$, with $J^3_L = J^3_R= - \frac{N}{2}$
\cite{Lerche:1989uy,Lerche:1989ig}.  Spectral flow takes the RR-state
$\ket{\psi_0}$ to the vacuum $\ket{1}_{\rm NS}$ of the NS sector.

\begin{figure}
\begin{quote}
  \begin{center}
  \begin{align*}
    \includegraphics[width=0.6\textwidth]{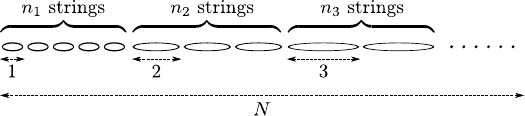} 
  \end{align*} 
 \end{center}
 \caption{\label{fig:eff_strings} 
 \sl The ``effective string'' picture of the
 RR ground states of the D1-D5 CFT\@.  There are $n_1$ strings of length
 $1$, $n_2$ strings of length $2$, and so on, and the total length of the system if $N$.}
\end{quote}
\end{figure}

The chiral primaries of the D1-D5 CFT can be obtained from the twist fields of the $S_N$ orbifold, and these fields are labeled by the conjugacy classes of $S_N$.  The
conjugacy classes of $S_N$ are in one-to-one correspondence with the partitions of $N$, which are given by  collections of non-negative
integers $\{n_{k}\}_{k\ge 1}$ satisfying
\begin{align}
 N  ~=~ \sum_{k\ge 1} k \, n_k.\label{partition_of_N}
\end{align}
It is useful to imagine these as describing a collection of ``effective strings.''  Namely, one associates the conjugacy class $\{n_k\}_{k\ge 1}$ with $n_1$ effective strings of length 1, $n_2$  effective strings of length 2, and so on.  The total length of all the effective strings is $N$.  See Fig.~\ref{fig:eff_strings}.
The effective string of length $k$ represents a twist field that intertwines $k$ copies of the $c=6$ CFT and may be viewed as taking $k$ circles of length $2 \pi R$ and twisting them into combinations of fewer but longer circles.
The maximally-spinning state $\ket{\psi_0}$ is unexcited by chiral primaries and so involves no intertwining of CFTs. It thus corresponds to the partition with $n_1=N$ and all other $n_k=0$.  

The holographic dual of the maximally-spinning state is a single, maximally-spinning, perfectly circular supertube in an $\IR^2$ plane.  In the near-supertube limit this geometry is exactly \emph{global} $AdS_3 \times S^3$.   The chiral primaries carry $\cR$-symmetry, by definition, and also have  $T^4$ indices.  In the effective string picture, we may view the effective strings as carrying  $\cR$-symmetry and $T^4$ indices coming from fermion zero modes.  We will focus here on the $\cR$-charge since it is visible from six-dimensional space-time and we will suppress for now the $T^4$ structure\footnote{For a more detailed description of the  geometries dual to effective strings that carrying $T^4$ indices, see \cite{Kanitscheider:2007wq}.}.  The partition \eqref{partition_of_N} is now refined according to
\begin{align}
 N ~=~  \sum_{k\ge 1}\sum_{\alpha,\dot{\alpha}=\pm}k \, n_k^{\alpha\dot{\alpha}},
 \label{partition_of_N_R-charge}
\end{align}
where $n_k^{\alpha\dot{\alpha}}=0,1,2,\dots$ is the number of effective strings with length $k$ and $SU(2)_L\times SU(2)_R$ spin $(\alpha,\dot{\alpha})$.  The maximally-spinning state $\ket{\psi_0}$ with $J^3_L = J^3_R= - \frac{N}{2}$ corresponds to the partition with $n_1^{--}=N$ and all other $n_k^{\alpha\dot{\alpha}}=0$.

Introducing twist fields generates excitations in the shape and density
modes and the bulk geometry dual to a generic two-charge state of the
form \eqref{partition_of_N_R-charge} is the Lunin-Mathur geometry
\cite{Lunin:2001jy} which is D1-D5 supertube with KKM
dipole charge and an arbitrary profile, or shape. (For a more detailed
dictionary see \cite{Kanitscheider:2006zf}.)  The Lunin-Mathur geometry
is completely regular \cite{ Lunin:2002iz} and parametrized by arbitrary functions of one
variable, $f^i(w)$ ($i=1,2,3,4$), describing the profile of the D1-D5
supertube in the $\bbR^4$ transverse to the D1-D5 world-volume. The
$SO(4)$ vector index $i$ of the $f^i(w)$ in $\IR^4$ is simply a pair of
spinor indices, $(\alpha,\dot{\alpha})$, of the $SU(2)_L\times SU(2)_R$
$\cR$-symmetry.  Hence we will denote these shape modes by
$f^{\alpha\dot{\alpha}}(w)$.  These functions are periodic,
$f^{\alpha\dot{\alpha}}(w+L)=f^{\alpha\dot{\alpha}}(w)$, and can be
expanded in Fourier series as
\begin{align}
 f^{\alpha\dot{\alpha}}(w)=\mu \sum_{\substack{k\in\bbZ\\ k\neq 0}}
 {a^{\alpha\dot{\alpha}}_k\over \sqrt{|k|}}e^{2\pi i k w/L},
 \qquad
  a^{\alpha\dot{\alpha}}_{-k}=(a^{\alpha\dot{\alpha}}_k)^*,
\end{align}
where $L,\mu$ are constants\footnote{Although we do not need their
explicit expression, for completeness, they are given by $L=2\pi g_s
\alpha' N_5/R$, $\mu=\alpha'^2 g_s/(R\sqrt{V_4})$, where $(2\pi)^4V_4$
is the volume of $T^4$ \cite{Mathur:2005zp}.}.  The zero mode $k=0$ has
been removed by shifting the origin of the $\bbR^4$.
The AdS/CFT dictionary for the two-charge states  \cite{Lunin:2001jy}\footnote{For a precise dictionary and its subtleties, see \cite{Kanitscheider:2006zf}.} is that the number of effective strings, specified by
$n^{\alpha\dot{\alpha}}_k$, is identified with the magnitude of the Fourier coefficients of the profile functions, $a^{\alpha\dot{\alpha}}_k$, by
\begin{align}
n^{\alpha\dot{\alpha}}_k 
 ~ \leftrightarrow ~
 |a^{\alpha\dot{\alpha}}_k|^2.\label{n_and_a}
\end{align}
In the bulk viewpoint, the constraint \eqref{partition_of_N_R-charge} is nothing other than the requirement that the supertube carries $N_1$ units of D1-brane charge.    

In this way one can substantiate the idea that semi-classical quantization of the D1-D5 profiles yields a description of the states of the D1-D5 system \cite{Lunin:2001jy, Rychkov:2005ji}.  For the two-charge system, the  profiles for the  typical  states have curvatures of order the Planck scale and so one must appeal to the idea of {\em microstate solutions} \cite{Bena:2013dka} discussed in the Introduction, to argue that while the supergravity approximation is not strictly valid, supergravity is capturing the essential semi-classical degrees of freedom that underlie the microstate structure. On the other hand, adding the third charge to the system means that there can be \emph{deep scaling} solutions  \cite{Bena:2006kb, Denef:2007vg, Bena:2007qc} in which the underlying structures remain macroscopic but lie at the bottom of long $AdS$ throats.  This means that  the supergravity approximation can remain valid over a large range of excitations and that the  semi-classical description of smooth low-curvature geometries may be enough to account for the entropy.

This dictionary \eqref{n_and_a} is in complete  accord with the idea that the effective strings carry $SU(2)_L\times SU(2)_R$ charges and they must represent visible microstates in the dual six-dimensional spacetime.  As we argued above, the effective strings arise from  twist fields that intertwine $k$ copies of CFT, with $k=1,\dots,N$.  The fact that these fields carry $\cR$-charges, {\it i.e.}, space-time angular momenta, means that they have polarizations directed into the space-time and so describe fluctuations in space-time.  Indeed, acting with these twist fields changes the length and spins of effective strings and, by the AdS/CFT dictionary \eqref{n_and_a}, corresponds to changing the shape of the back-reacted D1-D5 supertube.  We may look on these twist fields as providing a Landau-Ginzburg description of the shape modes of the D1-D5 system.  It should be stressed that these shape modes correspond to supertube profiles in the $\IR^4$ transverse to the D1-D5 world-volume.  There will be similar shape modes in the $T^4$ directions but in this paper we focus on the space-time shape modes.  

The correspondence between the quantization of shape modes and the states of the two-charge system is, of course, obvious in the F1-P duality frame where one is simply describing shape modes of a fundamental string.  Indeed, one can go from the F1-P modes to  the description of the D1-D5 modes by a suitable set of duality transformations. However, we need to work in the D1-D5 frame and  see that the states in this frame are also represented by shape modes  because we are now going to add a third charge to the system and it is easiest to understand what this entails if the new third charge is a momentum charge and not some other brane charge.  By showing that the D1-D5 states involve shapes as a function of one variable we are now going to see that the D1-D5-P states  are obtained by giving these D1-D5 shape modes an extra dependence on another direction.

\section{Adding momentum: the three-charge states}
\label{MomStates}

\subsection{Adding the momentum}
\label{MomAdd}

As we have seen, the two-charge ($\frac{1}{4}$-BPS) states of the D1-D5
system can be mapped onto the RR ground states of the CFT on the common
$S^1$ of the D1 and D5 branes.  The three-charge ($\frac{1}{8}$-BPS)
states are obtained simply if we keep the Ramond ground states in the
right-moving sector, thereby preserving half of the right-moving
supersymmetries, but allow any excited state, $\ket{\chi}$, in the
left-moving sector, thereby breaking all the left-moving
supersymmetries.  (The choice of the left/right sector to break/preserve
supersymmetry is purely conventional and we could have done it in the
other way around.)  The eigenvalue of the left-moving Virasoro
generator, $L_0$, on a state, $\ket{\chi}$, yields the momentum, $P=L_0-c/24$, of
the corresponding $\frac{1}{8}$-BPS state. It was this construction that
originally led to the perturbative counting of BPS microstates
\cite{Strominger:1996sh} and the microscopic description of the entropy
\eqref{count1}. As we saw above, the $\frac{1}{4}$-BPS shape modes along
the profile in the spatial $\IR^4$ are the shapes of the D1-D5
configuration described by $f^{\alpha\dot{\alpha}}(w)$ (or equivalently
by $a_k^{\alpha\dot{\alpha}}$) and these may be thought of as choices of
Ramond ground states or as the states generated by acting with chiral
primaries upon the maximally-spinning ground state $\ket{\psi_0}$.


Just as for fundamental strings, adding momentum to any system of
branes is expected to involve excitations transverse to the branes (see
footnote \ref{ftnt:transv_fluct}).  In the fully back-reacted
supergravity solution, these momentum states are reflected in a
non-trivial profile that sources the solution.  Conversely, the
quantization of that profile yields a semi-classical description of the
momentum states of the system.
If we assume that these are also true in the current situation, adding
momentum to the D1-D5 system means that the back-reacted supergravity
solution will now not only have a profile in the spatial $\IR^4$,
parametrized by $w$, but that such a profile will now also depend upon
$v$, the coordinate along the $S^1$ common to the D1 and the D5 branes\footnote{In general, the geometries
dual to CFT states that are exact eigenstates of the momentum operator
$P \equiv L_0-\tilde{L}_0$ are $v$-independent, while coherent states,
which are not a precise eigenstate of $P$, are $v$-dependent
\cite{Giusto:2013bda}.  We are concerned with the latter because we are
interested in the traveling waves on the supertube along $v$ and their
classical description is given by coherent states.}  Thus one obtains
shape modes that depend upon functions of \emph{two} variables and these
functions will provide a semi-classical description of all the states of
the D1-D5 system.

In particular, if we focus on the perturbative states visible within the space-time and described by $\cH_{\rm st}$ then  these shape modes are captured by the space-time shape modes of a generic, single superstratum.  We therefore expect that the two-charge profile functions, $f^{\alpha\dot{\alpha}}(w)$, which describe the supertube along an arbitrary curve in $\IR^4$, will be promoted to three-charge profile functions, $f^{\alpha\dot{\alpha}}(w,v)$, which describe the superstratum along an arbitrary surface.  Correspondingly, the one-index Fourier coefficients $a_{k}^{\alpha\dot{\alpha}}$ will be promoted to two-index ones, $a_{km}^{\alpha\dot{\alpha}}$.

Put differently, we can take a Landau-Ginzburg perspective in which the D1-D5 modes are created by chiral primaries and these, considered as Landau-Ginzburg fields, become momentum carriers simply through their descendant states within the left-moving Hilbert space.  Thus we see how a generic perturbative BPS excitation can give rise to a double Fourier series (with coefficients $a_{km}^{\alpha\dot{\alpha}}$) of space-time dependent excitations of the original D1-D5 system, or unexcited superstratum.

\subsection{Details of the perturbative momentum states}
\label{PertMomStates}

The connection between  perturbative CFT states and the supergravity shape modes can be made very explicit.  In the near-superstratum limit the geometry is simply $AdS_3 \times S^3$, which is the dual of the maximally-rotating RR ground state.  The shape modes of the superstratum are simply Fourier modes of supergravity fields on the $S^3$ and thus correspond to representations of   the $SU(2)_L\times SU(2)_R$.   While the two-charge D1-D5 shape modes carry quantum numbers of both $SU(2)_L$ and $SU(2)_R$, the momentum-carrying \emph{BPS} operators that excite those states carry only the quantum numbers of $SU(2)_L$ and hence adding momentum does not involve changing the D1-D5 shape modes that transform under $SU(2)_R$.  In particular, consider the maximally-spinning D1-D5 solution whose near-brane geometry is $AdS_3 \times S^3$.  The generic D1-D5 ground states can be thought of as fluctuation modes on the $S^3$.  In the NS sector, they are the chiral primary states and have quantum numbers under $SU(2)_L \times SU(2)_R$ given by $(\ell, m; \tilde \ell, \tilde m) = (\ell , \ell ; \tilde \ell, \tilde \ell)$.  Note that these D1-D5 ``supertube'' shape modes on the $S^3$ are very special, in that the quantum numbers are constrained to satisfy $\ell=m$, $\tilde{\ell}=\tilde{m}$ and, furthermore, $|\ell-\tilde{\ell}|$  is equal to the spin of the fields that exist in the theory.  For a fixed spin field the Fourier modes are determined by one quantum number and hence correspond to one-dimensional shape modes on the $S^3$.
In contrast, the BPS momentum carrying modes, which are of the form $(any,chiral)$ in the NS sector, allow more general excitations under $SU(2)_L$, while  the $SU(2)_R$ quantum numbers remain unchanged. So, the generic $\frac{1}{8}$-BPS mode will have $SU(2)_L \times SU(2)_R$ quantum numbers $(\ell, m; \tilde \ell, \tilde \ell)$. Since we now have $m$ independent of $\ell$, these will generate intrinsically two-dimensional shape modes on the $S^3$.

A particular subset of the BPS states involve arbitrary excitations created by operators in the $SU(2)$ current algebras, $J_{(r)}^{\alpha \beta}$, defined in \eqref{Jdefn}.  As noted above, these currents and the associated left-moving CFT in \eqref{ShapeCFT} reflect purely space-time modes and will be visible in the perturbative space-time shape modes of the superstratum. 

To make this more precise, one can easily describe the complete set of two-charge
supertube shape and density modes within supergravity and express the
result in terms of exact supergravity solutions in six dimensions.  One
can also realize the action of the superconformal algebra on the
geometry and, in particular, implement the action of the currents
\eqref{CFTUones} in terms of rotations on the supergravity solutions.
In this way one can, at the linearized level, generate the linearized
supergravity solutions with shape modes in the $(\ell, m; \tilde \ell,
\tilde \ell)$ representations by starting with the D1-D5 shape modes
$(\ell , \ell ; \tilde \ell, \tilde \ell)$ that correspond to chiral
primaries in CFT\@. Realizing this procedure has been one of the major
goals of \cite{Giusto:2012jx, Giusto:2013bda, Shigemori:2013lta}.  The fact that BPS
equations of the six-dimensional supergravity are essentially linear
means that knowing the linearized solutions is almost enough to
construct the fully back-reacted solutions \cite{Bena:2011dd}.  This
observation was exploited to significant effect in
\cite{Niehoff:2013kia, Shigemori:2013lta}.
To construct the fully back-reacted BPS fluctuations of the superstratum
and show that there is indeed an intrinsically two-dimensional BPS shape
modes in space-time one simply needs to take the special fluctuating
modes considered in \cite{Niehoff:2013kia} and use the current algebra
action, as in \cite{Mathur:2003hj, Shigemori:2013lta}, to find the
generic supergravity modes and then try to compute the fully
back-reacted solution using \cite{Bena:2011dd}.  

The foregoing procedure of rotating supertube fluctuation modes by the
generators of the asymptotic symmetry algebra corresponds to acting by
the \emph{total} $J^{\alpha\beta}=\sum_{r=1}^N J^{\alpha\beta}_{(r)}$
and not by the \emph{individual} $J^{\alpha\beta}_{(r)}$.  Moreover, one
really only needs the zero modes of $J^{\alpha\beta}$ to obtain the
fluctuations with quantum numbers of the form $(\ell, m; \tilde \ell,
\tilde \ell)$.  Put differently, this is equivalent to a rather trivial
statement that acting on a chiral primary by the generators of the
finite Lie algebras $SL(2,\bbR)_L\times SU(2)_L$ only gives the
descendant of a chiral primary but certainly does not yield generic
$\frac{1}{8}$-BPS states that are descendants of the non-chiral
primaries.  It therefore seems, at first sight, that the procedure we
have outlined only generates an extremely small subset of the general
momentum-carrying states, which require all the modes of all the
individual currents $J^{\alpha\beta}_{(r)}$.

However, this is not exactly what we are doing: we are not simply
rotating a complete, known classical BPS state.  Instead we are using
rotations to generate all the individual fluctuating modes of some of
the fields but discarding all of the rest of the rotated solution.  We
then take arbitrary linear combinations of those modes as seeds to generate new
classical solutions using the linear BPS system replete
with its sources that depend non-linearly on the fluctuating modes. In
this way we construct the most general, fully back-reacted fluctuating
supergravity solution.  In the quantum theory, classical solutions can
be regarded as coherent quantum states and so taking such  classical linear
combinations amounts to taking tensor products of the corresponding
quantum states.  The products of descendants of chiral primaries
generically yield the descendant of non-chiral primaries
\cite{deBoer:1998ip, deBoer:1998us}. Therefore, if we complete the fully
back-reacted supergravity solution based on linear combinations of modes, they will represent
the descendants of the non-chiral
primaries.

Thus the process of feeding a general superposition of classical
fluctuations into the complete BPS system will certainly generate the
most general exact classical BPS states and we claim that this will also
give a semi-classical description of the most general BPS quantum state.
Indeed, precisely this sort of result was established in
\cite{deBoer:1998us} where it was shown that the space of supergravity
fluctuations in a finite neighborhood of the $AdS_3\times S^3$
background precisely reproduced the elliptic genus of the CFT (Ref.\
\cite{deBoer:1998us} is when the internal manifold is K3; for $T^4$, see
\cite{Maldacena:1999bp}).

It is important to note that the result of
\cite{deBoer:1998us, Maldacena:1999bp} was only established using a
perturbative supergravity ``gas'' around a solution that lay outside the
black-hole regime and so one may quite reasonably doubt the
applicability of this result within microstate geometries that look like
black holes.  However, to make a microstate geometry that looks like a
black hole one does not simply use small perturbations of $AdS_3\times
S^3$: one must incorporate the back-reaction of the momentum to obtain
\emph{deep, scaling} microstate geometries in which the topological
cycles descend a long $AdS_2$ throat. We will discuss this further in
the next section, but here we want to note that $AdS_3\times S^3$
represents a good local model of individual topological bubbles and it
is expected that their fluctuations will give the microstate structure
only when these bubbles are located at the bottom of a \emph{deep,
scaling} throat.  All we therefore need from \cite{deBoer:1998us, Maldacena:1999bp} is the
result that the that semi-classical quantization of supergravity modes
on $AdS_3\times S^3$ captures the quantum CFT states locally.  It is
then expected that these states generate the correct microstate
structure of a black hole when they are located deep within a scaling
solution and greatly red-shifted as a result.

Before concluding this section we want to return to the other classical
modes that live on the internal $T^4$ and whose semi-classical
quantization should give rise to $\cH_{\rm int}$ in (\ref{HilbSpaces}).
Indeed, one of the points emphasized in \cite{Giusto:2012jx,
Giusto:2013bda} is that \emph{all} the perturbative excitations of D1-D5
system will be visible within the ten-dimensional supergravity
description of the superstratum. The left-moving $c=N$ theory
\eqref{ShapeCFT} whose states lie in $\cH_{\rm st}$ will indeed be
visible within the space-time of the effective six-dimensional theory
but the remaining modes, lying in $\cH_{\rm int}$ and described in terms
of the other $c=5 N$ part of the full CFT, will be also visible as
perturbative fluctuations of geometry and fluxes in the full
ten-dimensional solution. Thus, even though the space-time shape modes
of the superstratum will only lead to an entropy \eqref{count2}, one
might hope that the internal supergravity modes should lead to the full
accounting for the entropy \eqref{count1}.

However, as we will now describe, there is a subtlety in the
supergravity back-reaction that suggests that only the space-time shape
modes will have sufficient resolution to capture a large enough section
of the Hilbert space of the D1-D5-P system.

\subsection{The supergravity back-reaction and holography}
\label{SugrMomStates}

One of the important features of the CFT dual of black-hole microstates
is the fact that the CFT can have an energy gap as low as $E_{\rm gap}
\sim c^{-1} \sim \frac{1}{N_1 N_5}$.  This can be viewed as coming from
the scaling dimensions of the longest twist operators or from the
longest-wavelength momentum excitations of the longest effective
strings.  For a long time it was a puzzle as to how such fractionation,
and the energy gap in particular, could emerge from fluctuations of
smooth microstate geometries.  Such a match is crucial if the
semi-classical quantization of supergravity is to reproduce the
perturbative states of the CFT with sufficient fidelity to obtain the
entropy.

To understand the holographic description of the correct $E_{\rm gap}$,
one should first recall that the only way to construct microstate
geometries whose charges correspond to a five-dimensional black hole
with a finite horizon area is to use \emph{deep, scaling} BPS geometries
have a very long $AdS$ throat that is smoothly capped off by bubbles, or
homology cycles. The energy gap of these solutions then emerges
holographically \cite{Bena:2006kb} by taking the longest-wavelength
fluctuation of the microstate geometry and red-shifting it according to
the depth of the throat.  The depth of the throat is typically a free
classical parameter in the microstate geometry however semi-classical
quantization of such geometries sets the throat depth and thus fixes the
energy gap \cite{Bena:2007qc, deBoer:2008zn, deBoer:2009un}.  It was
thus one of the triumphs of the microstate geometry program that this
correctly reproduced the energy gap of the dual CFT\@.  The simplest
microstate geometries, in which the holographic energy gap was first
computed, can then be viewed as containing unexcited superstrata and so
the semi-classical quantization of the superstratum will reproduce the
correct energy levels.

Thus, in the holographic dual, modes of with energy $E_{\rm gap} \sim
\frac{1}{N_1 N_5}$ come from \emph{space-time} fluctuations whose
wavelengths are of order the diameter of throat of the BPS black
hole\footnote{This should, of course be defined as the area of the
throat to some suitable power.  Alternatively, for a microstate geometry
where the throat is capped off, this scale can also be defined by the
diameter of all the microstate structure.}.  If there is only a handful
of bubbles or superstrata, then this wavelength is set by the longest
wavelength fluctuation of homology cycles that spread across the throat.
If there are a lot of bubbles or superstrata then this wavelength should
be thought of as the longest wavelength collective mode of all the
bubbles and superstrata.

This result relies upon the crucial structure of the warp factors in the
metric.  In the IIB formulation, the ten-dimensional metric takes the
form:
\begin{align}
ds_{10}^2 & ~=~   - 2  \frac{1}{\sqrt{Z_1Z_2}} \, (dv+\beta) \big(du +  k  -  \coeff{1}{2}\,  Z_3\, (dv+\beta)\big) ~+~  \sqrt{Z_1Z_2} \, ds_4^2   ~+~   \sqrt{\frac{Z_1}{Z_2}} \,ds^2_{T^4} \nonumber \\
& ~=~  - \frac{1}{Z_3\sqrt{Z_1Z_2}}\,(dt+k)^2 ~+~  \frac{Z_3}{\sqrt{Z_1Z_2}}\,(dz+A^{(3)})^2  ~+~  \sqrt{Z_1Z_2}\, ds^2_4 ~+~  \sqrt{\frac{Z_1}{Z_2}} \,ds^2_{T^4} 
\,.   \label{tenmet}
\end{align}
For BPS solutions, the base metric, $ ds_4^2$, is hyper-K\"ahler and
ambi-polar; the deep, scaling solutions come from taking limits in which
a cluster of two-cycles in this base appear to scale to zero size. In
the physical metric \eqref{tenmet} the warp factor $(Z_1
Z_2)^{\frac{1}{2}}$ modifies this so that the cluster of cycles limits
to a finite size determined by $Q_1 Q_2$ in the spatial directions of
the base.  In the full ten-dimensional metric, the two-cycles are lifted
to three-cycles via the $v$ fiber and their volume also involves $Q_3$.
The important point is that the ``area'' of the throat scales with
$Q^{3/2}$ and so, as a result of the warp factor, the longest wavelength
mode that fits across the throat scales as $Q^{-1/2}$.  The red-shift of
the deep throat then gives an additional factor of $Q^{-3/2}$ to obtain
$E_{\rm gap} \sim Q^{-2}$ \cite{Bena:2006kb}.  On the other hand the
warp factors in the $T^4$ directions are $\cO(Q^0) = \cO(1)$ and so the
$T^4$ \emph{does not} expand to the typical size of the throat.  This
suggests that fluctuations around the $T^4$ will develop the wrong
energy gap, $E_{\rm T^4 \, gap} \sim Q^{-3/2}$.

Thus it seems that the supergravity fluctuations of the superstratum in
the space-time directions do give rise to the correct spectrum of
microstates but the supergravity fluctuations on the $T^4$ will lead to a
rather coarse sampling of the microstate structure.  It is possible that
our supergravity analysis of the $T^4$ fluctuations is too simplistic
and we will return to these issues in Section \ref{5/6} where we will
conjecture how the $T^4$ modes may ultimately be accounted for in the
supergravity back-reaction.

\subsection{Recapitulation}
\label{PertConclusions}

To finish this rather conservative analysis based upon perturbation
theory, we want to reiterate two important conclusions from our
discussion.  First, and most important, is that whatever the ultimate
outcome is on the holography of the $T^4$ modes, we have provided a good
match between the supergravity shape modes and the perturbative
microstate structure at least for the states in $\cH_{\rm st}$, with
central charge $c = N_1 N_5$.  Thus quantizing the superstratum should,
at least, reproduce \eqref{count2} and thus obtain the correct growth in
entropy with $N_1 N_5 N_P$.  This is already huge progress.
In particular, since these microstate geometries describe a
\emph{macroscopic} fraction of the black-hole entropy, this means that
all the typical states that contribute to the black-hole entropy will
have a finite transverse size.  Hence the entire system will not be
surrounded by a horizon and thus we will have established the fuzzball
proposal for BPS black holes in string theory.

The other thing we want to stress is that we have studied the
perturbative properties of a single, round superstratum and our work and
conclusions so far are based upon this rather conservative but fairly
detailed correspondence.  In Section \ref{5/6} and Section
\ref{Sect:Multi} we will argue that superstrata that have more
complicated shapes, and possibly split into bound states of multiple
superstrata will in fact be able to capture the full black-hole entropy.

\section{Towards the full black-hole entropy}
\label{5/6}

Our conservative counting of superstrata entropy in Section
\ref{MomStates} was based on the description of the maximally-spinning
supertube in the dual D1-D5 CFT and on the fact that in this CFT the
left-moving (supersymmetric) fermions are charged under $SU(2)_L$ but do
not carry $SU(2)_R$ angular momentum, and hence only a fraction of the
shape modes of the supertube will be able to carry momentum.  In this
section, we will be slightly bolder and discuss how the ``missing''
shape modes might re-emerge and account for the full entropy of the
D1-D5-P black hole.

\subsection{The shape modes of the superstratum}
\label{shapes/caveats}

From the perspective of the original argument for the existence of the
superstratum \cite{Bena:2011uw} and from the perspective of supergravity
solutions that describe certain superstratum components \cite{Bena:2011dd, Niehoff:2012wu}, the restriction
on the possible shape modes encountered in Section \ref{PertMomStates}
appears rather puzzling.

Indeed, if one constructs the superstratum by gluing together
16-supercharge plaquettes that preserve the D1-D5-P Killing spinors
irrespective or their orientation \cite{Bena:2011uw}, there appears to be no restriction on
the possible shapes of the resulting object, and hence the general
superstratum solution might be expected to be determined by four
arbitrary continuous functions of two variables.

This picture is further supported by the explicit construction of
supersymmetric solutions that have all the charges and dipole charges of
superstrata except one (the KKM dipole moment), and depend also on four
arbitrary continuous functions of two variables \cite{Niehoff:2012wu}.
These solutions are dubbed \emph{supersheets}.
Recall that, as mentioned in Section \ref{ss:superstrata}, the
first way to get a superstratum is to use a supertube transition to
``puff out'' D1 branes and momentum into a D1-P supertube and D5 branes and
momentum into a D5-P supertube (first stage), and then to use a second
supertube transition to puff out again the resulting (boosted and
rotated) D1-D5 system into a superstratum with KKM dipole charge (second stage).
Because supersheets do not have a KKM dipole moment, they must be
describing the first stage of this bubbling process and, consequently,
represent singular supergravity solutions.  The solution is expected to become a
smooth superstratum once the KKM dipole moment is added and it was shown
in \cite{Bena:2011uw} that adding the KKM dipole is compatible with
supersymmetry.  If the circle wrapped by the KKM dipole charge is small,
this will only affect the solution in the immediate vicinity of the
supersheets and hence one might reasonably expect that the KKM will not
upset the shape and the supersymmetry.

Based on the foregoing arguments, we are going to assume in the rest of
Section \ref{5/6} that a suitably generic superstratum can be given four
independent shape functions.  However, before proceeding on this
assumption, we wish to raise several issues that might lead to
restrictions on the BPS shape modes and limit such modes to those
described in Section \ref{MomStates}.

First, it was  noted in \cite{Bena:2011uw} that adding a KKM monopole requires
the orientation of the KKM to be properly aligned with the underlying
compactification circles, a fact that also was manifest in
\cite{Niehoff:2013kia} and leads, potentially, to restrictions on the orientations of
the solutions. Nevertheless, it is unclear whether this condition leads to significant restrictions
on the moduli space.  

Another issue is that the shape modes outlined in \cite{Bena:2011uw} were based
upon brane configurations that were not fully back-reacted and the description of shape modes was
based upon the local geometry of the solution.  In the  fully back-reacted 
superstratum some of the directions necessarily pinch off to make the smooth underlying topological cycles.   
Moreover, the directions that get pinched off  are typically those upon which the shape modes depend.  For 
a smooth solution the shape modes must therefore be required to die off as they approach these ``pinch-off'' points. 
This may well lead to restrictions on the allowed BPS modes that can be smoothly excited on a superstratum
and some of these restrictions were encountered and analyzed in \cite{Niehoff:2013kia}.  It remains to be
seen what the full range of allowable smooth shape modes can be for a single cycle but it may be only the
modes considered in Section \ref{MomStates}.

Finally, there is an interesting intermediate ground between the two
extremes of four shape modes and the modes of Section \ref{MomStates}.
It is possible that some of the shape modes have been suppressed by
focusing on a single topological cycle and, in particular, on the
scale-invariant $AdS_3 \times S^3$ near-superstratum limit.  The
``missing'' degrees of freedom could then emerge either as one restores
the asymptotic flatness or adds more structure so as to introduce a
scale. In the same vein, it may be that when one tries to make a KKM
resolution of a BPS supersheet of arbitrary shape, it is possible that
one may not be able to do it with a single topological bubble but that
it will require several such bubbles and that the combination of the
modes on such a multi-bubble solution can lead to more functions of two
variables.  We will pursue this idea further in
Section~\ref{Sect:Multi}.

\subsection{The MSW counting of black-hole entropy}

As we have argued, it is possible that once the full non-perturbative superstratum is constructed, the original picture of the 
BPS superstratum \cite{Bena:2011uw} could prove correct in terms of predicting the number of shape modes.
We will therefore examine what this would mean for the superstratum and in particular we will argue that 
that such fluctuation modes  reproduce \emph{all} the entropy of the
three-charge black hole.

To see how this comes about, it is useful to recall the ``second'' way to get a
superstratum by starting with a D1-D5 supertube with KKM dipole charge
and subsequently adding momentum to it.  Then the counting is very
similar to the Maldacena-Strominger-Witten (MSW) counting of the entropy
of four-dimensional black holes \cite{Maldacena:1997de}: One argues that
the number of momentum carriers on a superstratum is equal to the
dimension of the moduli space of deformations of the D1-D5 supertubes
and then derives the entropy by counting the ways of distributing the
momentum amongst these moduli.  At first glance the number of supertube
moduli is infinite, since an arbitrary shape can be decomposed into an
infinite Fourier series with arbitrary components. However, the
quantization of the shapes of the supertubes reduces the range of the
Fourier modes and hence renders the dimension finite. As we explained in
Section \ref{Sect:D1D5CFT}, this can be seen from the dictionary to the
dual D1-D5 CFT, which restricts the length of the maximal effective
string on the boundary (which corresponds to the Fourier mode of the
round supertube) to $N_1 N_5$, and since there are four functions
determining the embedding of the supertube in spacetime this corresponds
to a moduli space dimension $4 N_1 N_5$.\footnote{More precisely,
because of the constraint \eqref{partition_of_N_R-charge} imposed on the
$4N_1N_5$ Fourier modes, the moduli space dimension is $4 N_1 N_5-1$,
but this difference is negligible for the entropy counting.}

There is another way to figure out that the dimension of the moduli
space of spacetime deformations of two-charge supertubes is $4 N_1
N_5$. As we explained in Section \ref{Sect:2Chg}, these supertubes can
be dualized to fundamental strings carrying momentum, and the entropy of
this system comes from the various ways of splitting a given amount of
momentum, $N_P$, among different fractionated momentum carriers that
carry momentum quantized in units of $1/N_1$  \cite{Sen:1995in,Lunin:2001fv}. This entropy is given by
the number of possible ways of writing
\begin{equation}
N_1 N_P ~=~  \sum_{k\ge 1} k n_k  \,,
\end{equation}
much as in equation \eqref{partition_of_N}.  Upon taking into account
the fact that the fundamental string has eight species of bosonic
momentum carriers (corresponding to its 8 transverse directions) and
their fermionic partners, the number of partitions reproduces the
entropy of the two charge system. The dimension of the moduli space of
these configurations is given by the number of modes carrying momentum
that can be excited, and for one species alone this number is given by
the maximal value of $k$, which is the product of its two charges: $N_1
N_P$. Hence, the dimension of the moduli space of oscillations that will
become D1-D5 supertube oscillations in the {\em transverse}
four-dimensional space is again $4 N_1 N_5$.

One can also argue that the dimension of the supertube moduli space is
of order $N_1 N_5$ by considering the maximally-spinning (round)
supertube and counting its entropy \`a la Marolf and Palmer
\cite{Palmer:2004gu, Bak:2004rj, Bak:2004kz}.  This supertube has
angular momentum $J = N_1 N_5$, and if one tries to change its shape the
angular momentum becomes smaller. One can use the Born-Infeld action
describing this supertube to quantize the possible deformations of the
maximally-spinning supertube and find that this entropy comes from
integer partitions of $N_1 N_5 - J $. This counting therefore implies
that the dimension of the moduli space of a supertube with angular
momentum $J$ is equal to $N_1 N_5 - J$ (again for each bosonic
mode). Strictly speaking, this counting is only valid in the vicinity of
the maximally-spinning supertube configuration (when $N_1 N_5 - J \ll
N_1 N_5$), but if one extrapolates it to a supertube with zero angular
momentum one finds again the dimension of the moduli space of transverse
oscillations to be $4 N_1 N_5$.

In the foregoing discussion, we only counted the dimension of the moduli
space of the supertube fluctuations in the transverse non-compact
$\bbR^4$ directions (label them $1234$) and not the internal $T^4$
directions (label them $6789$). This restriction can be justified by a
supersymmetry analysis similar to the one in \cite{Bena:2011uw}.  As
mentioned above, the ``first'' way to get a superstratum is to first
puff out D1 branes and momentum, P, into a D1-P supertube inside
$\bbR^4$ and, simultaneously, puff out D5 branes and P into a D5-P
supertube inside $\bbR^4$. If the resulting D1-profile lies entirely
within the D5-profile, it is locally the same as the D1-D5 system which
can be puffed out again into a KKM dipole charge.  However, at the first
stage, instead of puffing out the D1 branes and P into a curve inside
$\bbR^4_{1234}$, we could have puffed them out into a curve inside
$T^4_{6789}$. For example, D1(5) and P(5) can be puffed out into D1(6)
and P(6) dipoles, where the numbers in the parentheses denote the
directions along which the object is extending.  Correspondingly,
D5(56789) and P(5) can be puffed out into D3(789) and F1(6) dipoles
(dissolved as fluxes inside the D5 worldvolume).  However, it is an
straightforward algebraic exercise \cite{Bena:2011uw} to show that these
puffed-out charges cannot undergo a second supertube transition.  Therefore,
interestingly, the second supertube transition is kinematically
(supersymmetrically) allowed only if the first transition is in the
transverse $\bbR^4$ directions. This holds true even if the internal
manifold is not $T^4$ but K3, because there is no difference between
$T^4$ and K3 in the local geometry.

Hence, the dimension of the moduli space of bosonic fluctuations of
D1-D5 supertubes in the transverse space is $4 N_1 N_5$. Much as for the
MSW black-hole entropy calculation, this dimension gives the number of
bosonic modes that carry momentum, and one expects by supersymmetry that
there should be an equal number of fermionic momentum carriers. As we
explained above, there is a tension between the perturbative analysis of
these modes (described in Section \ref{MomStates}) which indicates that
only $N_1 N_5$ of these modes can carry momentum supersymmetrically, and
the original argument for the existence of superstrata and the solutions
of \cite{Bena:2011dd, Niehoff:2012wu}, which suggests that all the four bosonic
modes, and hence all their four fermionic partners as well, can carry
momentum supersymmetrically.

If there really are four bosonic modes and four fermionic counterparts
then they will give a semi-classical description of momentum-carrying
states with $c= 6 N_1 N_5$, and the entropy of the superstrata is given
by the possible ways of carrying $N_P$ units of momentum:
\begin{equation}
S_{\rm superstrata} = 2 \pi \sqrt{{c \over 6} N_P}=  2 \pi \sqrt{N_1 N_5 N_P}\,,
\label{entropy}
\end{equation}
which reproduces exactly the Bekenstein-Hawking entropy of the
three-charge black hole. Since this entropy comes entirely from
spacetime modes and their fermionic partners, this entropy count also
reproduces the entropy of the D1-D5-P black hole if one replaces the
$T^4$ by $K3$.

We have thus argued that the shape modes of the superstratum have the
\emph{capacity} to describe a full set of semi-classical microstates of
a black hole and while this would represent a very happy state of
affairs, there are some words of caution to be made. First, as we
explained at the end of Section \ref{shapes/caveats}, adding a KKM
monopole and pinching off circles to make  topological cycles could potentially restrict
the shape modes \cite{Bena:2011uw}.  Second, we have argued that one
should think of the $4 N_1 N_5$ spatial shape modes of the superstratum
as independent ``moduli'' just as those of the MSW string and hence can
independently be assigned momentum states.  It remains unclear whether
these moduli are sufficiently independent and unobstructed. Indeed,
these excitations have to satisfy the constraint
(\ref{partition_of_N_R-charge}) and this restricts the size and
degeneracies of the putative moduli space.  This constraint will be
modified once one adds momentum and previously indistinguishable CFT
states become distinguishable.  Thus the independence of, and
restrictions upon, the supertube moduli remain unclear but as we have
seen, it is conceivable that the complete set of shape modes can capture
the complete BPS black-hole entropy.

\subsection{In search of the lost 5/6$^{\bf th}$'s}

The analysis of Section \ref{MomStates} starts from a single round
supertube, corresponding to a state of the D1-D5 CFT in which the long
effective string of length $N_1 N_5$ is split into $N_1 N_5$ effective
strings each of length one, and considers adding supersymmetric
(left-moving) momentum perturbatively on this object. The left-moving
momentum modes are only charged under $SU(2)_L$ but not under $SU(2)_R$,
which implies that only the modes that give one sixth of the central
charge of all the modes that one might have hoped to promote to momentum
carriers are in fact supersymmetric. Moreover, in the original
discussion of the superstratum \cite{Bena:2011uw} it was pointed out
that, while it seemed plausible that the shape modes could be excited
independently in the two directions of the superstratum surface, this
independence was not established rigorously.  So the most conservative
conclusion of the perturbative analysis of Section \ref{MomStates} is
that the space-time modes of superstrata are still given by functions of
two variables, as argued in \cite{Bena:2011uw}, but that these modes
only give $1\over \sqrt{6}$ of the entropy of the black hole.

It is important to examine the tension between the results of Section
\ref{MomStates} and the arguments of the previous subsection.  Indeed,
the results of Section \ref{MomStates} indicate that $5/6$ of the modes
that give rise to the black-hole entropy should appear as semi-classical
fluctuations on the internal $T^4$ and only $1/6$ of these modes are
visible in space-time.  This suggests that we should simply be looking
at the full supergravity solution in ten dimensions and the shape modes
on the $T^4$ in particular.  On the other hand the arguments we
presented above suggest that {\em all} the modes that carry the black
hole entropy can be visible as superstratum space-time modes.  We thus
appear to be in danger of over-counting.

One possible solution to this tension could be that the restrictions on
the supersymmetric momentum carriers coming from the perturbative
analysis are valid only in the vicinity of the maximally-spinning
supertube configuration in the free orbifold limit, and that far away
from that point in the CFT moduli space these restrictions will be
lifted.\footnote{Recall that the perturbation taking the CFT away from
the free orbifold point is a twist operator insertion which mixes
effective strings with different lengths.} Indeed, the supersheets of
\cite{Niehoff:2012wu} and other singular solutions that
have black hole charges and carry momentum with both $SU(2)_L$ and
$SU(2)_R$ angular momentum \cite{Bena:2011dd} can be thought of as limits of superstrata
solutions in which one has turned off the KKM dipole charge. This can be
done by making the radius of the second supertube transition very small,
which can be achieved by taking the number of KKM's to be very
large\footnote{This can appear paradoxical, but increasing the number of
KKM's decreases the radius of the KKM's and therefore reduces their
influence on the geometry}. From the perspective of the dual CFT, the
number of KKM's is the length of the effective strings, and increasing
this number brings one very far away from the state we considered in
Section \ref{MomStates}, where there are $N$ length-one effective
strings carrying $J_R^3=\cO(N)$ as a whole, towards the sector where
there are a few long effective strings of length $\cO(N)$ carrying
$J_R^3=\cO(1)$. Incidentally, this is also the sector where the
black-hole entropy lives, so if the superstratum counting that gives the
entropy (\ref{entropy}) is correct, this entropy comes exactly from
where it should come.  Starting with this sector with $J_R^3=\cO(1)$,
one has a large degree of freedom to increase/decrease $J_R^3$ by
creating short effective strings and making them carry the desired
$J_R^3$.  However, we must note that we do have the unitarity bound
$-{N\over 2}\le J_R^3\le {N\over 2}$, which is still in a apparent
conflict with the fact that, on the original supersheet, we could
consider arbitrary $SU(2)_R$ fluctuations.

Another possible way to reconcile the two analyses above could be to
consider multiple superstrata and allow different superstrata (or even
different parts of one superstratum) to have different orientations so
that the correlation with angular momentum might change between
superstrata.
It is possible for the momentum modes on one of
these superstrata to be charged under $SU(2)_L$ and for the modes on the
other to be charged under $SU(2)_R$. Thus, from a suitable distance, a
generic collection of superstrata could appear to replicate generic
space-time shape modes.  Moreover, it is possible to bring two
superstrata close to each other and to join them into a figure-eight
configuration that looks like a deformation of a superstratum with
dipole charge two. One can similarly argue that a superstratum with a
very large dipole charge, of the type that is expected to describe the
CFT states that give the black-hole entropy, can be deformed into
configurations that contain multiple superstrata, which can in turn
carry momentum modes with all angular momenta.

While these observations suggest that superstrata may have a much larger
set of space-time configurations than the single, round superstratum
considered in Section \ref{MomStates}, it does not resolve the
over-counting danger associated with having both the $T^4$ modes and the
full set of space-time shapes corresponding to states.  However, one can
argue that, in the regime of parameters where the black hole exists, the
modes that look like internal shape modes in the perturbative analysis
of Section \ref{MomStates} will be suppressed and, in addition, it is
possible that they give rise to fluctuations in the transverse space.

Indeed, our analysis of Section \ref{SugrMomStates} indicates that in
the fully back-reacted supergravity regime where the classical
black-hole solution exists, the modes that correspond to fluctuations in
the internal directions will have the wrong mass gap and will not be
therefore capable of describing the modes that give the black hole
entropy.  This will then suppress such semi-classical states in the
total entropy. A ``pessimist'' would then take the view that only the
perturbative space-time shapes have the correct energy gap and thus
contribute to the entropy, leading to the result (\ref{count2}).

However, based upon our experience with five-dimensional microstate
geometries, we know that details of ``internal sectors'' of the dual
field theory corresponding to degrees of freedom on the compactification
directions can become visible within the space-time geometry.  The
Coulomb-Higgs map \cite{Berkooz:1999iz,Bena:2012hf} is a classic example
in which Higgs-branch fields create composite operators that give rise
to strong effects within the space-time geometry that are more typically
associated with the Coulomb branch of the field theory.  Sometimes this
leakage of information onto the Coulomb branch can be complete in that
it yields complete information about the Higgs branch states and
sometimes it can be very incomplete in that it only captures a small
fraction the data about the ``internal states'' of the system.  Thus one
can take the optimistic view that the analysis of Section
\ref{SugrMomStates} suppresses the shape modes from exploring the $T^4$,
thereby protecting us from over counting, but these modes then leak into
the ``floppier'' space-time directions for which the energy gap is
\emph{much} lower.

It is also possible that the ``missing $5/6^{\rm th}$'s'' will not be
visible semi-classically within supergravity and that we can only obtain
the entropy (\ref{count2}).  As we have already stressed, this still
represents major progress.  On the other hand, we prefer to take the
optimistic view that the missing $5/6^{\rm th}$'s should still be
visible within supergravity.  One might therefore hope that the internal
shape modes of the single superstratum migrate to Coulomb branch and
become visible as space-time shape modes. It is interesting to ask whether these
modes will manifest themselves as superstratum modes, or as some other
mode complicated collective modes. The first possibility would reconcile
the superstratum analysis in this section with that of Section
\ref{MomStates}.  The second possibility would indicate there exists a
space-time object more complicated than the single, isolated
superstratum and such an object will account for $5/6$ of the modes that give the
entropy of a black hole, while the single, isolated superstratum
accounts for the other $1/6$. This more complicated object might be some
multi-superstrata state or even something new.  Either way, finding and
understanding this more complicated object would clearly be a key
priority.

We now make some first steps in suggesting the role of multi-superstrata
states.

\section{Multi-superstrata}
\label{Sect:Multi}


Independent of the bulk considerations of the previous section, we will argue
that the structure of the three-charge states in CFT suggests that bound states of multiple
superstrata are the most natural candidate for the holographic duals of the CFT states. To
explain this, we begin by unpacking more of the details of the states
described in Sections {\ref{Sect:2Chg}~and~\ref{MomStates}}.

\subsection{Structure of three-charge states in CFT}

In the D1-D5 CFT, a two-charge BPS state, {\it i.e.}\ the RR ground
state is made of multiple effective strings of various length.  Ignoring
the $SU(2)_L\times SU(2)_R$ charge, it is specified by the numbers
$\{n_k\}_{k\ge 1}$ satisfying \eqref{partition_of_N} and is of the
following form:
\begin{align}
 \prod_{k\ge 1} (\ket{0}_k)^{n_k}
 =
 (\ket{0}_1)^{n_1}
 (\ket{0}_2)^{n_2}
 (\ket{0}_3)^{n_3}
 \cdots,\label{2-chg_state}
\end{align}
where $\ket{0}_k$ is the ground state of the $c=6k$ CFT living on the
effective string of length $k$.  See Fig.\ \ref{fig:eff_strings}. The
bulk dual of this is a D1-D5 supertube whose profile function $f(w)$ has
Fourier coefficients $a_k$ given by
\begin{align}
 |a_1|^2=n_1,\qquad
 |a_2|^2=n_2,\qquad
 |a_3|^2=n_3,\qquad \dots.
\end{align}
Note that we are ignoring the $SU(2)_L\times SU(2)_R$ charge for simplicity of presentation and therefore the spin indices $\alpha,\dot{\alpha}$ on $f(w),a_k$ are also omitted.

\begin{figure}
\begin{quote}
  \begin{center}
  \begin{align*}
    \includegraphics[height=0.30\textwidth]{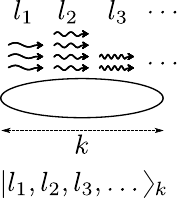} 
  \end{align*} 
 \end{center}
 \caption{\label{fig:single_string_excited} 
 \sl The excited state
 $\ket{l_1,l_2,\dots}_k$ of a single effective string of length $k$. On
 the string, we have $l_1$ quanta carrying $1\over k$ units of momentum,
 $l_2$ quanta carrying $2\over k$ units of momentum, and so on.
 The standard projection in the orbifold procedure imposes the condition $\sum_k m l_m/k\in \bbZ$.
\vspace*{-.5cm}}
\end{quote}
\end{figure}
The three-charge states are obtained by exciting momentum-carrying modes
on the effective strings.  In particular, on an effective string of
length $k$ lives the $SU(2)_L$ current $J^3_L(z)$,\footnote{Here,
$J^3_L(z)$ is defined to be $J^3_L(z)=J^{3}_{L(r)}(z)$ with
$2\pi(r-1)\le \arg(z) <2\pi r$ and is multi-valued, where $r=1,\dots,k$
is the copy index.  In particular, $J^3_L(z)$ is not the sum of the individual
currents, $\sum_{r=1}^k J^3_{L(r)}(z)$.  } whose modes we denote by
$J_{m\over k}$, $m\in\bbZ$.  Note that the mode numbers are in units of
${1\over k}$ because the length of the string is $k$.  We
can use these modes to obtain momentum-carrying states on a single
effective string as follows:
\begin{align}
\label{momexcit}
 (J_{-{1\over k}})^{l_1}(J_{-{2\over k}})^{l_2}\cdots
 \ket{0}_k\equiv \ket{l_1,l_2,\dots}_k,
\end{align}
with the $S_N$-orbifold constraint that the total momentum on the
effective string is an integer, namely, $\sum_{m\ge 1} {ml_m/ k}\in
\bbZ$.  See Fig.\ \ref{fig:single_string_excited} for a pictorial
description of this state.  Since the modes $J_{-{m\over k}}$ carry
non-vanishing $SU(2)_L$ charge, they are visible in six-dimensional
space-time.  If we excite the $J^3_L$ modes on all the effective strings
in the two-charge state \eqref{2-chg_state}, we obtain the general
three-charge state that can be created by $J^3_L$
excitations.\footnote{Of course, there are other momentum-carrying
states that cannot be obtained by the action of $J^3$ but, for
simplicity, we focus on the states that can be simply labeled as in
\eqref{momexcit}.} In doing so, we must remember that effective strings
of identical length $k$ are indistinguishable if they are in the ground
state but, once we excite $J^3_L$ modes, they become distinguishable
(unless they have identical excitation numbers $\{l_1,l_2,\cdots\}$).
Thus, for each $k$, the $n_k$ states will be broken into distinguishable
and indistinguishable effective strings.

To be concrete, let us focus on effective strings with one particular
value of length $k$, say, $k=3$, for a moment.  If we have, {\it e.g.},
seven of strings of length 3, we have the following two-charge state:
\begin{align}
 (\ket{0}_3)^7.\label{2chg_k=3}
\end{align}
The seven strings are indistinguishable because they are all in the same
ground state.  So, this  two-charge state is completely specified by
a single number $n_3=7$. Now, three-charge states are obtained by
exciting momentum modes on these strings, as in \eqref{momexcit}. For
example, take two of them and excite the first ($m=1$) momentum mode
three times on each; namely, we have two strings, all in the state
$(J_{-{1\over 3}})^3\ket{0}_3=\ket{3,0,0,\dots}_3$.  For four of the
remaining five strings, excite the $m=1$ mode once and the $m=2$ mode
four times; namely, all four strings are in the state $(J_{-{1\over
3}})(J_{-{2\over 3}})^4\ket{0}_3=\ket{1,4,0,\dots}_3$.  Finally, let the
last string be in the state $(J_{-{1\over 3}})^6(J_{-{3\over
3}})^1\ket{0}_3=\ket{6,0,1,\dots}_3$.  Note that the total momentum in
each string is an integer.  The three-charge state thus obtained is
\begin{align}
 (\ket{3,0,0,\dots}_3)^{2}\, (\ket{1,4,0,\dots}_3)^{4}\,
  (\ket{6,0,1,\dots}_3)^{1}\,.\label{3chg_k=3}
\end{align}
The $n_3=7$ indistinguishable strings in \eqref{2chg_k=3} have split
into three distinguishable groups.  If $n_3^{(i)}$ denotes the number of
strings in the $i^{\rm th}$ group, we have the splitting
\begin{align}
 n_3=7=2+4+1=\sum_{i=1}^3 n_3^{(i)}.\label{n_3constraint}
\end{align}
The $n_3^{(i)}$ strings in the $i^{\rm th}$ group are all in the same
excited state and indistinguishable.  Let $n_{3 m}^{(i)}$, $m\ge 1$
denote the momentum excitation numbers for the state of the $i^{\rm th}$
group. In the present example,
\begin{align}
\begin{split}
  \text{$1^{\rm st}$ group:} \quad (n_3^{(1)}\equiv n_{30}^{(1)};\,n_{31}^{(1)},\,n_{32}^{(1)},\,n_{33}^{(1)},\,\dots)
 =(2;3,0,0,\dots),\\
 \text{$2^{\rm nd}$ group:}  \quad (n_3^{(2)}\equiv n_{30}^{(2)};\,n_{31}^{(2)},\,n_{32}^{(2)},\,n_{33}^{(2)},\,\dots)
 =(4;1,4,0,\dots),\\
 \text{$3^{\rm rd}$ group:}  \quad (n_3^{(3)}\equiv n_{30}^{(3)};\,n_{31}^{(3)},\,n_{32}^{(3)},\,n_{33}^{(3)},\,\dots)
 =(1;6,0,1,\dots),\\
\end{split}
\label{n_3m(i)}
\end{align}
where we defined $n_{30}^{(i)}\equiv n_3^{(i)}$.  More generally, it is
clear that the general three-charge state of length-3 strings is
completely specified by the numbers $\{n_{3m}^{(i)}\}_{m\ge
0,i\ge 1}$.  Distinguishability between different groups with $i\neq i'$
means that $\{n_{3m}^{(i)}\}_{m\ge 1}\neq \{n_{3 m}^{(i')}\}_{m\ge 1}$.

The general three-charge state built on the general two-charge state
\eqref{2-chg_state} is obtained by multiplying excited strings with
different values of $k$ together.  Namely, for each $k$, we index the
distinguishable families of momentum excitations by $(i)$ and let $n_{k
0}^{(i)}$ denote the number of \emph{indistinguishable} strings in each
family (they are indistinguishable because they have identical
excitation numbers).  Therefore, the two-charge constraint
\eqref{partition_of_N} is refined to:
\begin{equation}
\sum_{i\ge 1} \, n_{k 0}^{(i)} ~=~ n_k \,, \qquad \sum_{k\ge 1} \sum_{i\ge 1} \, n_{k 0}^{(i)} ~=~ N \,.\label{nk0(i)_sum_rule}
\end{equation}
Let $n_{k m}^{(i)}$ ($m\ge 1$) denote the momentum excitations, as in
\eqref{n_3m(i)}, of the $i^{\rm th}$ set of effective strings of length
$k$:
\begin{equation}
 \ket{n_{k 1}^{(i)} \,, n_{k 2}^{(i)}\,,\cdots}_k \,.
 \end{equation}
Distinguishability from the other strings of length $k$ means that the
momentum excitations must be different: $\{n_{k m}^{(i)}\}_{m\ge 1}\neq
\{n_{k m}^{(i')}\}_{m\ge 1}$ if $i\neq i'$.

The three-charge states thus obtained are:
\begin{align}
 \prod_{k\ge 1} \prod_{i\ge 1}
 \left(\ket{n_{k1}^{(i)},n_{k2}^{(i)},\dots}_k\right)^{n_{k0}^{(i)}}
 &=
 \left(\ket{n_{11}^{(1)},n_{12}^{(1)},\dots}_1\right)^{n_{10}^{(1)}}
 \left(\ket{n_{11}^{(2)},n_{12}^{(2)},\dots}_1\right)^{n_{10}^{(2)}}
 \cdots
 \notag\\
 &\qquad\times
 \left(\ket{n_{21}^{(1)},n_{22}^{(1)},\dots}_2\right)^{n_{20}^{(1)}}
 \left(\ket{n_{21}^{(2)},n_{22}^{(2)},\dots}_2\right)^{n_{20}^{(2)}}
 \cdots\label{3chg_state_CFT}
\end{align}
where the powers represent the fact that there are $n_{k0}^{(i)}$
indistinguishable effective strings in the same state.  The three-charge
states \eqref{3chg_state_CFT} are thus specified by the non-negative
integers, $\{n_{k m}^{(i)}\}$.  The index $k\ge 1$ is associated with
the Fourier mode in the $w$-direction (the loop in $\IR^4$ of the
original D1-D5 system) and the index $m\ge 0$ is associated with the
momentum Fourier modes in the $v$-direction.  Note that we have
identified $n_{k 0}^{(i)}$ introduced above \eqref{nk0(i)_sum_rule} with
the $m=0$ mode number.  Thus we have sufficient data to describe the
shape modes as a function of two variables, as expected of a
superstratum.  However, there remains an additional index $(i)$ --- this
means that \emph{the general three-charge states in the D1-D5 CFT
naturally parametrize multiple functions of two variables}.  What is
the physical interpretation of this fact?

\subsection{Multi-superstrata interpretation}

\begin{figure}[htbp]
\begin{quote}
  \begin{center}
  \begin{align*}
    \includegraphics[width=.9\textwidth]{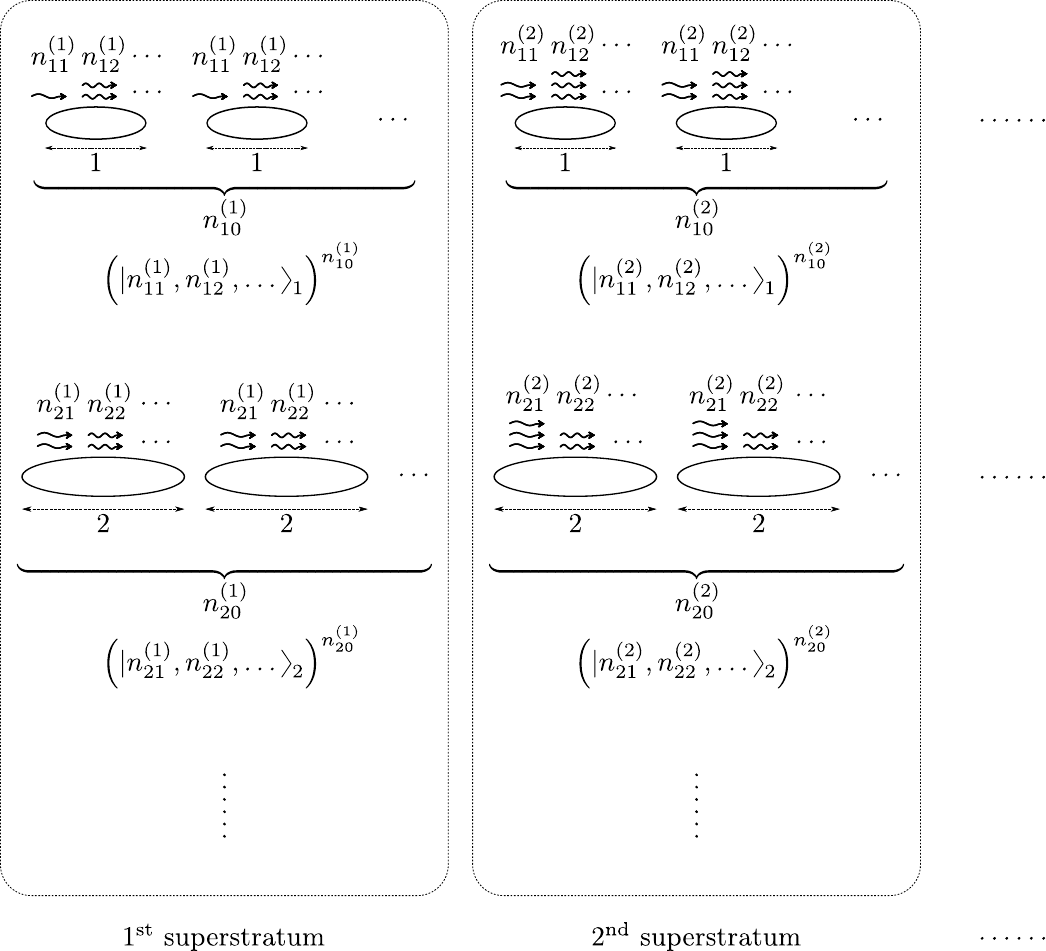} 
  \end{align*} 
 \end{center}
 \caption{\label{fig:multiple_strings_excited} \sl Momentum carrying
 excitations on multiple effective strings and their possible
 multi-superstratum interpretation.  For each string length $k$, strings
 on which identical momentum modes are excited are grouped together.
 For fixed $k$, the $n_{k 0}^{(1)}$ strings in group 1 are all in the
 same state $\ket{n_{k1}^{(1)},n_{k2}^{(1)},\dots}_k$ and are
 indistinguishable, the $n_{k 0}^{(2)}$ strings in group 2 are all in
 the same state $\ket{n_{k1}^{(2)},n_{k2}^{(2)},\dots}_k$ and are
 indistinguishable, and so on.  The shape of the 1\textsuperscript{st}
 superstratum is specified by the number of strings in group 1 for all
 possible values of $k$, namely by $\{n_{k m}^{(1)}\}$.  The shape of
 the 2\textsuperscript{nd} superstratum is specified by $\{n_{k m}^{(2)}\}$, and so
 on.  See the text for more detail.}
\end{quote}
\end{figure}

The index $(i)$ labels distinguishable effective strings of the same
length: sets of effective strings that only became distinguishable by virtue of the
momentum excitations on them.  It is therefore tempting to interpret
$(i)$ as labeling the multiple superstrata into which the original D1-D5
supertube has split.  The momentum excitations promote the original
profile function, $f(w)$, into a function of two variables, $f(v, w)$,
but we conjecture that the two-charge profile function actually gets
promoted into multiple functions of two variables labeled by $(i)$:
\begin{align}
 f(w)
 \quad\to\quad
 f^{(1)}(w,v),\quad 
 f^{(2)}(w,v),\quad 
 f^{(3)}(w,v),\quad \dots\ ,\label{f_promotion}
\end{align}
where $f^{(i)}(w,v)$ describes the world-volume of the $i^{\rm th}$ superstratum.  The Fourier coefficients $a_{km}^{(i)}$ of these functions are then  given by
\begin{align}
 |a_{km}^{(i)}|^2=n_{km}^{(i)}.
\end{align}
See Fig.\ \ref{fig:multiple_strings_excited} for a pictorial description
of the state \eqref{3chg_state_CFT} and the multi-superstrata interpretation.

We hasten to note the important fact that the foregoing description of
three-charge states, such as \eqref{3chg_state_CFT}, is valid only at
the free orbifold point in the moduli space of the D1-D5 CFT, whereas
the actual supergravity sits at a very different point in the moduli
space.  Deforming the CFT away from the orbifold point corresponds to
turning on twist operator perturbations (see \cite{Avery:2010qw} for a
recent detailed account).  Twist operators mix different twist sectors
and therefore the picture of each individual state gets modified.
However, it is the \emph{number} of states that is important for our
proposal, and it is not changed by such deformations.  Namely, the
deformation does not change the crucial fact that more data than can fit
on a single superstratum is needed to account for general three-charge
states.  Therefore, this does not invalidate our proposal that general three-charge states are
represented by multiple superstrata,
although the precise dictionary between the superstrata shape functions
$f^{(i)}(w,v)$ and the CFT states may not be as simple as described
above.  For example, it is quite conceivable a state that looks like a
multi-strata state in CFT corresponds to a single-stratum state in
supergravity, and vice versa.\footnote{This point is particularly clear
for the three-charge states built on the two-charge state with
$n_1=N,n_{k\ge 2}=0$ (short string sector) which we denote by
$\ket{\psi_0}$. The foregoing CFT picture (the $k=1$ version of
\eqref{3chg_k=3}) says that we can build multiple-strata states on this
state.  On the other hand, in supergravity, the state $\ket{\psi_0}$
corresponds to a single, circular, unexcited superstratum, which is
nothing but pure $AdS_3\times S^3$.  Momentum-carrying excitations on it
are small deformations of the $S^3$, which do not seem to lead to
multi-superstrata.} This is analogous to the fact that, in
$AdS_5/CFT_4$, once interactions are turned on, the single/multi-trace
operator basis of the CFT Hilbert space is different from (and a unitary
transformation of) the single/multi-particle basis in of the
supergravity Hilbert space.

Our multi-superstrata proposal raises several important issues.  First,
all the states we are discussing in \eqref{3chg_state_CFT} are
states within \emph{the same CFT} and \emph{not} states in distinct
CFT's.  Arguing that some of these states correspond to different
superstrata suggests that we are factoring the CFT into different CFT's.
At a more basic level, if one accepts that the distinguishable families
factor into different superstrata then why do we not accept that the
same must happen in the two-charge D1-D5 system: Why aren't effective
strings of different lengths simply different supertubes?

The resolution of all these issues comes from remembering that multiple
supertubes have no $E \times B$ interactions, and therefore can be
separated at arbitrary distances. If we consider a solution that
contains only two-charge supertubes placed at the bottom of a long $AdS$
throat, these supertubes are not trapped at the bottom of the throat and
can move freely out of the throat. They represent therefore unbound
states dual to factorized CFT's. On the other hand, two generic
superstrata will always have non-trivial $E \times B$ interactions, and
hence a solution that has multiple superstrata at the bottom of a long
$AdS$ throat will represent a bound state of the CFT\@. Solutions with
different numbers of superstrata will have different topology, and hence
will belong to different sectors of this CFT\@.

Another important consideration is the fact that the bubbling transition
to create microstate geometries with non-trivial cycles requires the
three-charge system.  The bubble equations
\cite{Bates:2003vx,Bena:2004de,Gauntlett:2004qy}, which relate the sizes
of cycles to the fluxes through those cycles, degenerate for two charges
or if a flux through a cycle vanishes and so the corresponding bubble
collapses.  Thus the possibility of separate superstrata forming a bound
state in a CFT can only occur if one excites the momentum modes in the
D1-D5 system and only if one excites momenta in distinct ways so that
the fluxes on bubbles do not vanish.  Conversely, if two superstrata
have exactly the same shape and charge distribution then they will
coalesce within a given $AdS$ throat or, if they are not in an $AdS$
throat, there will be no force between them and they can be moved
arbitrarily far away from each other, which is not describable within
one dual CFT \cite{Seiberg:1999xz}.

It is worth noting that the ``moulting phase'' of the D1-D5 system
\cite{Bena:2011zw} that appears in the three-charge situation with large
angular momentum has structures rather similar to the ones proposed
here.  In \cite{Bena:2011zw}, the following problem was studied: for
given momentum charge and angular momentum $J_L=\cO(N)$, what is the
ensemble of states that has the largest entropy?
In the CFT (at the orbifold point), the most entropic states were found to
be made of two sectors of effective strings, reminiscent of
\eqref{f_promotion}.  The first sector is made of a long string with
length $\cO(N)$, which carries all the momentum charge as well as
the entropy, while the second sector consists of many ($\cO(N)$) short
strings of length one, which carry $J_L,J_R=\cO(N)$ but no entropy.
On the other hand, in supergravity, the most entropic configuration was
found to be a two-center solution in an asymptotically $AdS$ space.  One
center is a BMPV black hole carrying all the momentum charge and
entropy, while the other center is a supertube carrying $J_L,J_R=\cO(N)$
but no entropy.\footnote{Although the configurations in CFT and
supergravity seem quite similar to each other, the entropy of the CFT
states and that of the bulk two-center solution do not quite agree (the
CFT entropy is always larger than the supergravity entropy), which is
presumably caused by the partial lifting of states at strong coupling.}
(Because the BMPV black hole can be thought of as ``shedding'' or
``moulting'' a supertube, it was dubbed the ``moulting phase'').
The fact that the multi-sector states of the CFT correspond to a
multi-center solution in supergravity can be thought of as evidence in
support of our conjecture (even though these configurations are not
microstates but phases with finite entropy).

Apart from the natural way in which the correspondence of
distinguishable twisted sectors and bound states of multiple superstrata
appears to work, one can obtain further evidence for the conjecture by
re-examining the arguments of
\cite{Bena:2006kb,Bena:2007qc,deBoer:2008zn,deBoer:2009un} that obtain
the CFT gap from the supergravity solution.  We first note that the
longest effective string corresponds to
\begin{equation}
n_N ~=~ 1 \,, \qquad n_k  ~=~ 0 \,, \ \ 1\le k < N  \,,
\end{equation}
and so can only involve a single superstratum, no matter how we add
momentum.  This sector of the theory is also the sector with $E_{\rm
gap} \sim \frac{1}{N_1 N_5}$ and was obtained holographically by
considering an excitation of a bubbled geometry that has a wavelength
equal to the size of the $AdS$ throat.  Such a wavelength would be the
natural fundamental oscillation of a superstratum whose scale is that of
the entire throat.  In multiple, bound superstrata the bubbles of
geometry will be smaller than the throat and the scale of an individual
bubble will be roughly set by the scale of the throat divided by the
some appropriate power of the number of bubbles.  Thus the fundamental
modes of such individual bubbles will have a shorter wavelength and a
higher energy gap.  Indeed, the energy gap of such a configuration
should be $E_{\rm gap} \sim \frac{p}{N_1 N_5}$, where $p$ is the
approximate number of bubbles that span the ``diameter'' of the throat.
This, at least qualitatively, fits very nicely with the corresponding
decreased lengths of the effective strings in the CFT\@.  Obviously more
work is needed to fully substantiate our conjecture but we think it is
promising enough to warrant our description here.

\section{Conclusions}
\label{Sect:Concl}

In this paper we have argued that the BPS microstates of the D1-D5-P
system will manifest themselves in the regime in which the classical
black hole exists as smooth horizonless ``superstratum''
solutions. Despite the absence of an explicit solution describing the
generic superstratum, we have been able to account for their
entropy using the intuition that adding momentum modes to any system of
branes will, upon back-reaction, emerge as shape modes in supergravity,
and, conversely, that the semi-classical quantization of such shape
modes will reconstruct the original Hilbert space of momentum states.

We first considered the construction of a superstratum in terms of
fluctuations around a maximally-spinning supertube and have argued, from
the dual D1-D5 CFT, that the number of supersymmetric momentum carriers
of the superstratum is given by the product, $N_1 N_5$, of its D1 and D5
charges.  This conservative estimate, which we believe can be
substantiated with a high level of confidence, gives the entropy:
\begin{equation}
S ~=~ 2 \pi \sqrt{\frac{1}{6} N_1 N_5 N_P} \label{count2a} 
\end{equation}
and this is expected to come entirely from smooth supergravity solutions.

Then we went on to make a somewhat bolder proposal for counting the
entropy of superstrata using an approach similar to that of Maldacena,
Strominger and Witten \cite{Maldacena:1997de}.  Specifically, we argued
that the space of transverse fluctuations of two-charge supertubes must
have dimension $4N_1 N_5$.  One can then view this as the moduli space
of the superstratum and, much as in the original construction of
superstrata \cite{Bena:2011uw}, all these moduli could carry momentum.
Assuming these moduli are independent and unobstructed, there are thus
$4N_1 N_5$ bosonic modes which, when combined with their fermionic
superpartners, would give an entropy:
\begin{equation}
S ~=~ 2 \pi \sqrt{ N_1 N_5 N_P} \label{count1a} \,.
\end{equation}
This exactly matches the black-hole entropy. We have also discussed the
possible ways to reconcile this estimate to the more conservative
estimate above, and have argued that, in the regime of parameters where
the black hole exists, all the modes in the internal directions should
somehow manifest themselves as fluctuations in the transverse space. We
have also argued that one cannot match all the states of the CFT by
counting perturbatively around a single superstratum solution, and that
multiple superstrata bound states are a natural candidate for matching
these states.

Modulo the explicit construction of superstratum solutions that depend
on arbitrary functions, we have presented what we believe to be strong
evidence that the so-called fuzzball proposal is the correct description
of extremal supersymmetric black holes within string theory. Indeed, if one
can obtain a \emph{macroscopic} fraction of the black-hole entropy from
horizonless supergravity solutions, this implies that all the typical
states that contribute to the black-hole entropy will have a finite
transverse size, and hence the entire system will not be surrounded by
horizon.   This in turn would imply that the correct way to think
about the textbook black-hole solution is as a thermodynamic
approximation of a huge number of horizonless configurations, much as
a continuous fluid is a thermodynamic approximation of a huge number of
molecule configurations.

The conservative and bolder views of superstrata lead to significant differences 
in the structure of typical black-hole microstates.  If all the black-hole microstates 
are visible as transverse superstrata modes, then it is possible that upon full 
back-reaction these modes will all give rise to low-curvature solutions that 
have a long black-hole-like throat and end in a smooth cap. This would imply that 
the modes captured by six-dimensional supergravity are enough to account for the black-hole
entropy, which would establish the fuzzball proposal in its strong form.

If, however, only $1/\sqrt{6}$ of the black-hole entropy comes from
transverse modes, then the typical black-hole microstates will still be 
horizonless, but will not be describable as smooth solutions of
six-dimensional supergravity:  The typical microstates will necessarily involve stringy or
Kaluza-Klein modes.   This would establish the ``weak version'' of the
fuzzball proposal, which is enough for solving the information paradox,
but it may not offer us a framework, at least within supergravity, for doing rigorous 
computations that could help establish, for example, whether an incoming observer feels a
firewall or falls through the fuzzball states unharmed.

Clearly, there are two essential steps that should be done next.  The
first is the explicit construction of the superstratum solutions that
depend on functions of two variables. This would represent major progress 
toward establishing the fuzzball proposal for extremal black holes. 
The dramatic simplification of the BPS system of equations underlying these solutions
\cite{Bena:2011dd} means that it might be possible to construct the BPS
supergravity excitations at full non-linear order. The discussion at the
beginning of Section \ref{MomStates} showed that arbitrary space-time
shape modes break all the supersymmetry and that only the
representations $(\ell, m; \tilde \ell, \tilde \ell)$ of $SU(2)_L \times
SU(2)_R$ can be excited in the $\frac{1}{8}$-BPS superstratum. This
observation also underlies the analysis in \cite{Niehoff:2013kia,
Shigemori:2013lta} and it will provide invaluable insight into how to
address the construction of a fully back-reacted superstratum that
depends upon a general function of two variables.

The second, and most difficult, step is to extend this work to
non-extremal black holes.  A very useful insight comes of our analysis
here where we noted that certain momentum carriers that are charged
under $SU(2)_R$ may break supersymmetry\footnote{A similar phenomenon
happens for supertubes, and there taking into account the
supersymmetry-breaking modes is crucial if one is to quantize correctly
the supersymmetric modes \cite{Palmer:2004gu}.}.  Hence, adding these
fluctuations to a typical BPS superstratum state may allow us to move
away from extremality and to argue that the supergravity structure of
the black-hole microstates that we have analyzed in this paper is robust
when supersymmetry is broken. This, in turn, would imply that
near-extremal, and quite possibly generic, black holes are thermodynamic
approximations of horizonless solutions and that the pure states of a
black hole would be represented by horizonless configurations. This
would solve the black-hole information paradox and allow us to address,
far more rigorously, the puzzles that the information-theory analysis of
black hole has revealed \cite{Braunstein:2009my, Almheiri:2012rt,
Mathur:2012jk, Susskind:2012rm, Bena:2012zi, Susskind:2012uw,
Avery:2012tf, Avery:2013exa, Almheiri:2013hfa, 
Verlinde:2013uja, Mathur:2013gua, Harlow:2013tf}.

\section*{Acknowledgments}

We would like to thank Stefano Giusto and Rodolfo Russo for extremely
useful discussions that strongly influenced the work presented here.
MS thank the IPhT, CEA-Saclay for hospitality where part of
this work was done.
NPW is grateful to the IPhT, CEA-Saclay, the Institut des Hautes \'Etudes
Scientifiques (IH\'ES), Bures-sur-Yvette and the Yukawa Institute for
hospitality while various parts of this work was done.  NPW would also
like to thank the Simons Foundation for their support through a Simons
Fellowship in Theoretical Physics.
We are all grateful to the Centro de Ciencias de Benasque for
hospitality at the ``Gravity -- New perspectives from strings and higher
dimensions'' workshop, and to the Yukawa Institute at Kyoto University
for hospitality at the ``Exotic Structures of Spacetime'' workshop
(YITP-T-13-07).
The work of IB was supported
in part by the ERC Starting Independent Researcher Grant 240210-String-QCD-BH, by the
John Templeton Foundation Grant 48222: ``String Theory and the Anthropic Universe'' and
by a grant from the Foundational Questions Institute (FQXi) Fund, a donor advised fund of
the Silicon Valley Community Foundation on the basis of proposal FQXi-RFP3-1321 to the
Foundational Questions Institute.
The work of MS was supported in part by Grant-in-Aid for Young
Scientists (B) 24740159 from the Japan Society for the Promotion of
Science (JSPS)\@.
The work of NPW was supported in part by the DOE grant
DE-FG03-84ER-40168.



\end{document}